\documentclass[a4paper,fleqn,usenatbib]{mnras}

\usepackage{newtxtext,newtxmath}

\usepackage[T1]{fontenc}
\usepackage{ae,aecompl}
\usepackage{bm}

\usepackage{float}
\usepackage[export]{adjustbox}
\usepackage{graphicx}	

\usepackage{amsmath}	
\usepackage{amssymb}	

\usepackage{multicol}
\usepackage{verbatim}   
\usepackage{threeparttable}
\usepackage{upgreek}
\usepackage{rotating}

\usepackage{subfig}
\usepackage{comment}
\usepackage{longtable,lscape}
\usepackage{caption}
\usepackage{threeparttable}
\newcommand{\lcdm}{$\Lambda$CDM}

\newcommand{\om}{\Omega_{m0}}
\newcommand{\ol}{\Omega_{\Lambda}}
\newcommand{\ok}{\Omega_{k0}}
\newcommand{\FT}[1]{}

\title[Study of QSO data]{Determining the range of validity of quasar X-ray and UV flux measurements for constraining cosmological model parameters}

\author[]{
Narayan Khadka ,$^{1}$\thanks{E-mail: nkhadka@phys.ksu.edu}
Bharat Ratra,$^{1}$\thanks{E-mail: ratra@phys.ksu.edu}
\\
$^{1}$Department of Physics, Kansas State University, 116 Cardwell Hall, Manhattan, KS 66502, USA\\
}

\date{Accepted XXX. Received YYY; in original form ZZZ}

\pubyear{2019}

\begin{document}
\label{firstpage}
\pagerange{\pageref{firstpage}--\pageref{lastpage}}
\maketitle

\begin{abstract}
We use six different cosmological models to study the recently-released compilation of X-ray and UV flux measurements of 2038 quasars (QSOs) which span the redshift range $0.009 \leq z \leq 7.5413$. We find, for the full QSO data set, that the parameters of the X-ray and UV luminosities $L_X-L_{UV}$ relation used to standardized these QSOs depend on the cosmological model used to determine these parameters, i.e, it appears that the full QSO data set include QSOs that are not standardized and so cannot be used for the purpose of constraining cosmological parameters. Subsets of the QSO data, restricted to redshifts $z \lesssim 1.5-1.7$ obey the $L_X-L_{UV}$ relation in a cosmological-model-independent manner, and so can be used to constrain cosmological parameters. The cosmological constraints from these lower-$z$, smaller QSO data subsets are mostly consistent with, but significantly weaker than, those that follow from baryon acoustic oscillation and Hubble parameter measurements.
\end{abstract}

\begin{keywords}
\textit{(cosmology:)} cosmological parameters -- \textit{(cosmology:)} observations -- \textit{(cosmology:)} dark energy
\end{keywords}



\section{Introduction}
\label{sec:Introduction}
Observational astronomy has established that the universe is currently under going accelerated cosmological expansion \citep{Ratra2008,Martin2012, Ellis2020}. If general relativity is an accurate description of gravitation, to explain this observational fact we need dark energy. In general relativistic cosmological models dark energy contributes $\sim 70\%$ of the current energy budget. The simplest cosmological dark energy model is the spatially-flat $\Lambda$CDM model \citep{peeble1984}, which is consistent with most observational data \citep{Farooq2017, Scolnic2018, Plank2018, eBOSS2020}. In this model, spatial hypersurfeces are flat, the dark energy density is the spatially-homogeneous and time-independent cosmological constant ($\Lambda$), cold dark matter (CDM) contributes $\sim 25\%$ of the current energy budget, with baryonic matter contributing most of the remaining $\sim 5\%$.

While many different cosmological observations are consistent with  a constant $\Lambda$ and flat spatial hypersurfaces, these data do not yet strongly rule out mildly dynamical dark energy density or weakly curved spatial hypersurfaces\footnote{Discussion of observational constraints on spatial curvature may be traced through \cite{Farooq2015}, \cite{chen6}, \cite{Yu2016}, \cite{Rana2017}, \cite{Ooba2018a, Ooba2018b, Ooba2018c}, \cite{Yu2018}, \cite{Park2018a, Park2018b, Park2018d, Park2018c, Park2019}, \cite{wei2018}, \cite{DESa}, \cite{Cole2019}, \cite{Jesus2019}, \cite{Handley2019}, \cite{Zhai2019}, \cite{Li2019}, \cite{Geng2020}, \cite{Kumar2020}, \cite{Efsta2020}, \cite{Di2020}, \cite{Gao2020}, \cite{Abbassi2020}, \cite{Yang2020}, \cite{Ruiz2020}, \cite{Fabris2020}, \cite{Vagnozzi2020a, Vagnozzi2020b}, and references therein. These papers discuss constraints determined using various combinations of data, including baryon acoustic oscillation, Hubble parameter, cosmic microwave background anisotropy, supernova apparent magnitude, growth factor, and other measurements.}, so in this paper we consider cosmological models that incorporate these phenomena.

Two main goals of cosmology are to establish the most accurate cosmological model and to tighten the cosmological parameter constraints as much as possible. To accomplish these we should use all observational data. Observational data to date have largely been restricted to two parts of redshift space. More widely used low redshift observational data lie in the redshift range $0 \leq z \leq 2.3$, which include baryon acoustic oscillation (BAO) measurements, Type Ia supernova data, and Hubble parameter $[H(z)]$ observations, while higher redshift cosmic microwave background anisotropy data probe redshift space at $z \sim 1100$. 

In the intermediate redshift range $2.3 \leq z \leq 1100$ cosmological models are poorly tested. In this range there are a handful of data sets. These include HIIG starburst galaxy data that reach to $z \sim 2.4$ \citep{Siegel2005, Plionis2009, Mania2012, Charez2014, Gonzalez2019, Caoka, Caokb}, quasar angular size measurements that reach to $z \sim 2.7$ \citep{Gurvits1999, chen3a, Cao2017, Ryan2019, Caoka, Caokb}, and gamma-ray burst observations that reach to $z \sim 8.2$ \citep{Lamb2000, SamushiaR2010, Wang2016, Demianski2019, Amati2019, Dirisa2019, Marco2020, Luongo2020, Khadka2020c, Caokb}. Quasar (QSO) X-ray and UV flux measurements that reach to $z \sim 7.5$ provide another set of data that probe this intermediate redshift region \citep{Risaliti2015, Risaliti2019, yang2019, Velten2020, wei2020, Lin2019, Zheng2020, Mehrabi2020, Khadka2020a, Khadka2020b, Lusso2020,Rezaei2020, Speri2020}.

In this paper we study the new QSO compilation \citep{Lusso2020} containing 2421 QSO measurements (of which 2038 are of higher quality, which are what we use here). These QSOs reach to  $z \sim 7.5$ and are thought to be standardizable through a phenomenological relation between the QSO X-ray and UV luminosities, the $L_X-L_{UV}$ relation. In this paper, we examine these 2038 QSO measurements and investigate their reliability as standard candles. We find that for the full (2038) QSO data set, the $L_X-L_{UV}$ relation parameter values can depend significantly on the cosmological model used in the analysis of these data. This could mean that some of these quasars are not properly standardized and implies that the full QSO data set should not be used to constrain cosmological parameters.\footnote{There are some curious patterns in the differences between $L_X-L_{UV}$ relation parameters for pairs of cosmological models. See Sec.\ 5.1.} 

To examine this issue more carefully we separately analyze the low-redshift and high-redshift halves of the full QSO data set, containing 1019 QSOs at $z < 1.479$ and 1019 at $z > 1.479$. We find somewhat significantly different $L_X-L_{UV}$ relation parameters for the higher-$z$ and lower-$z$ data subsets for the same cosmological model and in the $z > 1.479$ data subset we find significant differences in the $L_X-L_{UV}$ relation parameters for different models. However, for the $z < 1.479$ data subset (and somewhat less so for the $z < 1.75$ QSO data subset) the $L_X-L_{UV}$ relation parameters are independent of cosmological model and so these smaller, lower-$z$, QSO data subsets appear to mostly contain standardized candles and so are suitable for the purpose of constraining cosmological parameters. 

While the $z < 1.479$ and $z < 1.75$ QSO data subsets favor larger current non-relativistic matter density parameter $(\Omega_{m0})$ values (as did the 2019 QSO data, see \cite{Khadka2020b}, and as do the full (2038) QSO data here), the uncertainties on $\Omega_{m0}$ are larger for these smaller QSO data subsets, so these $\Omega_{m0}$ values do not that significantly disagree with values determined using other cosmological probes. Additionally the cosmological parameter constraints from the $z < 1.479$ and $z < 1.75$ QSO data subsets are largely consistent with those from BAO + $H(z)$ data (unlike for the full (2038) QSO data constraints), so it is reasonable to jointly analyze a lower-$z$ QSO data subset with the BAO + $H(z)$ data. However, the statistical weight of the smaller lower-$z$ QSO data subsets are low compared to that of the BAO + $H(z)$ data and so adding the QSO data to the mix does not significantly alter the BAO + $H(z)$ data cosmological constraints.

In Sec. 2 we summarize the models we use in our analyses. In Sec. 3 we describe the data we test and use. In Sec. 4 we outline the techniques we use in our analyses. In Sec. 5 we present the results of the QSO consistency tests and the cosmological parameter constraints from all the data used in this paper. We conclude in Sec. 6.

\section{Models}
\label{sec:models}
In this paper we derive model parameter constraints by comparing observed cosmological quantities at known redshift with corresponding model predictions. To make predictions we use six different general relativistic dark energy cosmological models. Three consider flat spatial geometry, the other three consider spatially non-flat geometries. Fundamentally, the predictions in our analyses here are related to the cosmological expansion rate---the Hubble parameter---which is a function of redshift $z$ and the cosmological parameters of the model.

In compact form, the Hubble parameter in all six models can be written as
\begin{equation}
\label{eq:friedLCDM}
    H(z) = H_0\sqrt{\Omega_{m0}(1 + z)^3 + \Omega_{k0}(1 + z)^2 + \Omega_{DE}(z)},
\end{equation}
where $H_0$ is the Hubble constant, and $\Omega_{k0}$ is the current value of the curvature energy density parameter. For spatially-flat models $\Omega_{k0} = 0$. For analyses of the BAO + $H(z)$ and QSO + BAO + $H(z)$ data compilations, we split $\Omega_{m0}$ in terms of the current value of the baryon density parameter $(\Omega_{b})$ and the current value of the cold dark matter density parameter $(\Omega_{c})$ through the equation $\Omega_{m0}$ = $\Omega_{b}$ + $\Omega_{c}$. In four of the six models, $\Omega_{DE}(z)$ = $\Omega_{DE0}(1+z)^{1+\omega_X}$, where $\Omega_{DE0}$ is the current value of the dark energy density parameter and $\omega_X$ is the dark energy equation of state parameter (defined below).

In the $\Lambda$CDM model, $\omega_X = -1$, and $\Omega_{DE}$ = $\Omega_{DE0}$ = $\Omega_{\Lambda}$ is the cosmological constant dark energy density parameter and is a constant. The current values of the three energy density parameters are related by the energy budget equation, $\Omega_{m0} + \Omega_{k0} + \Omega_{\Lambda} = 1$. In the spatially-flat $\Lambda$CDM model it is conventional to choose $\Omega_{m0}$ and $H_0$ to be the free parameters while in the spatially non-flat $\Lambda$CDM model $\Omega_{m0}$, $\Omega_{k0}$, and $H_0$ are taken to be the free parameters. For analyses of the BAO + $H(z)$ and QSO + BAO + $H(z)$ data, instead of $\Omega_{m0}$ we use $\Omega_b h^2$ and $\Omega_c h^2$ as free parameters; here $h$ is the dimensionless Hubble constant defined by the equation $H_0$ = 100$h$ ${\rm km}\hspace{1mm}{\rm s}^{-1}{\rm Mpc}^{-1}$.

In the XCDM parametrization the dynamical dark energy $X$-fluid pressure $P_X$ and energy density $\rho_X$ are related through the equation of state $P_X = \omega_X \rho_X$ where $\omega_X$ is the equation of state parameter for the dynamical dark energy $X$-fluid, and $\Omega_{DE0}$ = $\Omega_{X0}$ is the current value of the $X$-fluid dynamical dark energy density parameter. The current values of the three energy density parameters are related as $\Omega_{m0} + \Omega_{k0} + \Omega_{X0} = 1$. When $0 > \omega_X > -1$, the $X$-fluid energy density decreases with time. In the spatially-flat XCDM parametrization it is conventional to choose $\Omega_{m0}$, $\omega_X$, and $H_0$ to be the free parameters while in the non-flat XCDM parametrization, $\Omega_{m0}$, $\Omega_{k0}$, $\omega_X$, and $H_0$ are chosen to be the free parameters. For the BAO + $H(z)$ and the QSO + BAO + $H(z)$ data analyses, instead of $\Omega_{m0}$, we use $\Omega_b h^2$ and $\Omega_c h^2$ as free parameters. In the limit $\omega_X = -1$ the XCDM parametrization reduces to the $\Lambda$CDM model.

In the $\phi$CDM model the dynamical dark energy is a scalar field $\phi$ \citep{peebles1988, Ratra1988, Pavlov2013}.\footnote{For constraints on the $\phi$CDM model determined using various different combinations of observational data see \cite{yashar2009}, \cite{Samushia2010}, \cite{camp}, \cite{Farooq2013b}, \cite{Farooq2013a}, \cite{Avsa}, \cite{Sola2017}, \cite{Sola2018, Sola2019}, \cite{Zhai2017}, \cite{Ooba2018b, Ooba2018d}, \cite{Sangwan2018}, \cite{Park2018a}, \cite{Singh2019}, \cite{Caokb}, \cite{Roy2020} and references therein.} In this model $\Omega_{DE}$ = $\Omega_{\phi}(z, \alpha)$, the scalar field dynamical dark energy density parameter, is determined by the scalar field potential energy density, for which we assume an inverse power law form,
\begin{equation}
\label{eq:phiCDMV}
    V(\phi) = \frac{1}{2}\kappa m_{p}^2 \phi^{-\alpha}.
\end{equation}
Here $m_{p}$ is the Planck mass, $\alpha$ is a positive parameter, and $\kappa$ is a constant whose value is determined using the shooting method to guarantee that the current energy budget equation $\Omega_{m0} + \Omega_{k0} + \Omega_{\phi}(z = 0, \alpha) = 1$ is satisfied.

With this potential energy density, the equations of motion of a spatially homogenous cosmological model are
\begin{equation}
\label{field}
    \ddot{\phi} + 3\frac{\dot{a}}{a}\dot\phi - \frac{1}{2}\alpha \kappa m_{p}^2 \phi^{-\alpha - 1} = 0,
\end{equation}
and,
\begin{equation}
\label{friedpCDM}
    \left(\frac{\dot{a}}{a}\right)^2 = \frac{8 \uppi}{3 m_{p}^2}\left(\rho_m + \rho_{\phi}\right) - \frac{k}{a^2}.
\end{equation}
Here overdots denote derivatives with respect to time, $k$ is positive, zero, and negative for closed, flat, and open spatial hypersurfaces (corresponding to $\Omega_{k0} < 0, =0, \rm and >0$), $\rho_m$ is the non-relativistic matter energy density, and $\rho_{\phi}$ is the scalar field energy density
\begin{equation}
    \rho_{\phi} = \frac{m^2_p}{32\pi}[\dot{\phi}^2 + \kappa m^2_p \phi^{-\alpha}].
\end{equation}
By solving eqs. (3) and (4) numerically we can compute $\rho_{\phi}$ and then compute $\Omega_{\phi}(z, \alpha)$ by using the equation
\begin{equation}
    \Omega_{\phi}(z, \alpha) = \frac{8 \uppi \rho_{\phi}}{3 m^2_p H^2_0}.
\end{equation}
In the spatially-flat $\phi$CDM model it is conventional to choose $\Omega_{m0}$, $\alpha$, and $H_0$ to be the free parameters and in the non-flat $\phi$CDM model, $\Omega_{m0}$, $\Omega_{k0}$, $\alpha$, and $H_0$ are taken to be the free parameters. For the BAO + $H(z)$ and QSO + BAO + $H(z)$ data analyses, instead of $\Omega_{m0}$ we use $\Omega_b h^2$ and $\Omega_c h^2$ as free parameters. In the limit $\alpha\rightarrow0$ the $\phi$CDM model reduces to the $\Lambda$CDM model.

\begin{figure}
    \includegraphics[width=\linewidth, right]{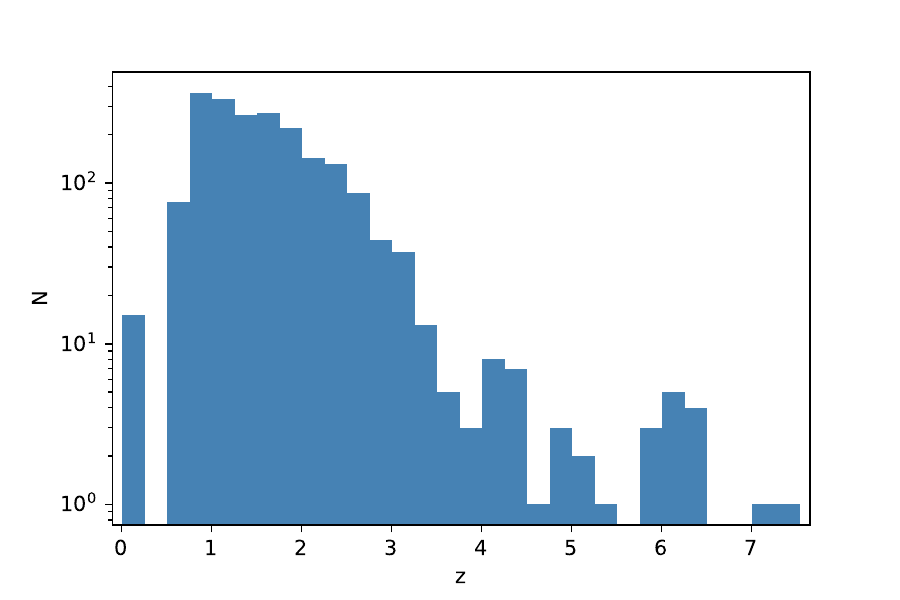}\par
\caption{Redshift distribution of the 2038 QSO that we use in our analyses.}
\label{fig: A Hubble diagram of quasar}
\end{figure}

\begin{table}
\small\addtolength{\tabcolsep}{-1.2pt}
\caption{BAO data. $D_M \left(r_{s,{\rm fid}}/r_s\right)$ and $D_V \left(r_{s,{\rm fid}}/r_s\right)$ have units of Mpc. $H(z)\left(r_s/r_{s,{\rm fid}}\right)$ has units of ${\rm km}\hspace{1mm}{\rm s}^{-1}{\rm Mpc}^{-1}$ and $r_s$ and $r_{s, {\rm fid}}$ have units of Mpc.}
\begin{tabular}{cccc}
\hline
$z$ & Measurement & Value  & Ref.\\
\hline
$0.38$ & $D_M\left(r_{s,{\rm fid}}/r_s\right)$ & 1512.39 & \cite{alam}\\
\hline
$0.38$ & $H(z)\left(r_s/r_{s,{\rm fid}}\right)$ & 81.2087 & \cite{alam}\\
\hline
$0.51$ & $D_M\left(r_{s,{\rm fid}}/r_s\right)$ & 1975.22 & \cite{alam}\\
\hline
$0.51$ & $H(z)\left(r_s/r_{s,{\rm fid}}\right)$ & 90.9029 & \cite{alam}\\
\hline
$0.61$ & $D_M\left(r_{s,{\rm fid}}/r_s\right)$ & 2306.68 & \cite{alam}\\
\hline
$0.61$ & $H(z)\left(r_s/r_{s,{\rm fid}}\right)$ & 98.9647 & \cite{alam}\\
\hline
$0.122$ & $D_V\left(r_{s,{\rm fid}}/r_s\right)$ & $539 \pm 17$ & \cite{carter2018}\\
\hline
$0.81$ & $D_A/r_s$ & $10.75 \pm 0.43$ & \cite{DESb1}\\
\hline
$1.52$ & $D_V\left(r_{s,{\rm fid}}/r_s\right)$ & $3843 \pm 147$ & \cite{ata2018}\\
\hline
$2.334$ & $D_M/r_s$ & 37.5 & \cite{du}\\
\hline
$2.334$ & $D_H/r_s$ & 8.99 & \cite{du}\\
\hline
\end{tabular}
\label{tab:BAO}
\end{table}


\section{Data}
\label{sec:data}
Since 2015, Lusso, Risaliti, and collaborators have worked on trying to standardize high redshift quasars by using the X-ray and UV fluxes of QSOs and the $L_X-L_{UV}$ relation \citep{Risaliti2015, Risaliti2019, Lusso2020}. We have previously studied the cosmological consequences of the 2015 and 2019 QSO data and noted that the 2019 QSO data favor a higher value for $\Omega_{m0}$ than do most other cosmological probes \citep{Khadka2020a, Khadka2020b}. 

The 2020 compilation \citep{Lusso2020} contains 2421 quasar measurements over the redshift range $0.009 \leq z \leq 7.5413$. In these data, the wanted 2500 {\AA} UV fluxes for some $z < 0.7$ QSOs are determined by extrapolation from the optical,  which is less reliable because of possible host-galaxy contamination. At $z < 0.7$, 15 local QSOs, whose 2500 {\AA} flux are determined from the UV spectra without extrapolation, provide higher quality data than the other $z < 0.7$ sources \citep{Lusso2020}. So in this paper we use 2038 quasar measurements, 2023 QSOs at $z > 0.7$ and the 15 higher quality ones at $z < 0.7$, to constrain the cosmological and $L_X-L_{UV}$ relation parameters in six different cosmological dark energy models. The redshift distribution of these data is shown in Fig.\ 1.

We find that the full 2038 QSO sample has $L_X-L_{UV}$ relation parameters that depend on the cosmological model used in the analysis and this suggests that some of the QSOs in the full QSO sample are not standard candles. To study this issue, we first divide the full QSO sample into two equal groups, a low redshift half with $z < 1.479$ (hereafter QSO$-z < 1.479$) and a high redshift half with $z > 1.479$ (hereafter QSO$-z > 1.479$), each containing 1019 QSO measurements. We chose this redshift because these two subsets contain an equal number of measurements. Additionally, $z \sim 1.5$ is close to the upper limit of redshift space that has been reasonably accurately probed by other data, such as SNIa, BAO, and $H(z)$ measurements and it is of interest to see whether the $z < 1.5$ QSO data constraints are consistent with those from the other measurements. We find that the $L_X-L_{UV}$ relation parameters determined from the QSO$-z < 1.479$ sample are independent of the cosmological model used in the analysis, however the QSO$-z > 1.479$ $L_X-L_{UV}$ relation parameters are quite model dependent. To estimate the highest redshift of QSOs in this compilation \citep{Lusso2020} that obey the $L_X-L_{UV}$ relation in a model-independent manner, and thus are potentially standardizable candles, we also consider three other QSO data subsamples with redshifts up to $z < 1.75$ (hereafter QSO$-z < 1.75$), $z < 2$ (hereafter QSO$-z < 2$), and $z < 2.25$ (hereafter QSO$-z < 2.25$). QSO$-z < 1.75$, QSO$-z < 2$, and QSO$-z < 2.25$ contain 1313, 1534, and 1680 QSO measurements, respectively.

In this paper we also use 11 BAO measurements spanning the redshift range $0.0106 \leq z \leq 2.33$ and 31  $H(z)$ measurements spanning the redshift range $0 \leq z \leq 1.965$. The BAO measurements are given in Table 1 of this paper and include the first 9 of Table 1 of \cite{Caoka} and the two new measurements of \cite{du} (see the following section for BAO covariance matrices), and the 31 $H(z)$ measurements are given in  Table 2 of \cite{Ryan2018}.

The QSO$-z < 1.479$ (and less so the QSO$-z < 1.75$) data have almost cosmological-model-independent $L_X-L_{UV}$ relation parameters and so seem to be potentially safely standardizable, and the QSO$-z < 1.479$ data constraints are quite consistent with those from the BAO + $H(z)$ data, so we also use the QSO$-z < 1.479(/1.75)$ data in combination with BAO + $H(z)$ data to constrain cosmological and $L_X-L_{UV}$ relation parameters.

\section{Methods}
\label{sec:methods}
Quasar X-ray and UV luminosities of selected quasars are known to be correlated through the non-linear $L_X-L_{UV}$ relation \citep{Risaliti2015, Risaliti2019, Khadka2020a, Khadka2020b}. This relation is
\begin{equation}
\label{eq:XCDM}
    \log(L_{X}) = \beta + \gamma \log(L_{UV}) ,
\end{equation}
where $\log$ = $\log_{10}$ and $\gamma$ and $\beta$ are free parameters to be determined from the data. For the 2015 and 2019 QSO compilations \citep{Risaliti2015, Risaliti2019}, the values of $\gamma$ and $\beta$ determined from the data are independent of the cosmological model used in the determination \citep{Khadka2020a,Khadka2020b}.

Luminosities and fluxes are related through the luminosity distance so eq.\ (7) can be rewritten as
\begin{equation}
\label{eq:XCDM}
    \log(F_{X}) = \beta +(\gamma - 1)\log(4\pi) + \gamma \log(F_{UV}) + 2(\gamma - 1)\log(D_L),
\end{equation}
where $F_{UV}$ and $F_X$ are the quasar UV and X-ray fluxes. Here $D_L(z, p)$ is the luminosity distance, which is a function of redshift $z$ and a set of cosmological parameters $p$ and is given by
\begin{equation}
\label{eq:DM}
  \frac{H_0\sqrt{\left|\Omega_{k0}\right|}D_L(z, p)}{(1+z)} = 
    \begin{cases}
    {\rm sinh}\left[g(z)\right] & \text{if}\ \Omega_{k0} > 0, \\
    \vspace{1mm}
    g(z) & \text{if}\ \Omega_{k0} = 0,\\
    \vspace{1mm}
    {\rm sin}\left[g(z)\right] & \text{if}\ \Omega_{k0} < 0,
    \end{cases}   
\end{equation}
where
\begin{equation}
\label{eq:XCDM}
   g(z) = H_0\sqrt{\left|\Omega_{k0}\right|}\int^z_0 \frac{dz'}{H(z')},
\end{equation}
and $H(z)$ is a Hubble parameter that is described in Sec.\ 2 for each model.

Given a cosmological model, eqs. (8) and (9) can be used to predict X-ray fluxes of quasars at known redshift. We compare these predicted fluxes with observations by using the likelihood function \citep{Khadka2020a}
\begin{equation}
\label{eq:chi2}
    \ln({\rm LF}) = -\frac{1}{2}\sum^{N}_{i = 1} \left[\frac{[\log(F^{\rm obs}_{X,i}) - \log(F^{\rm th}_{X,i})]^2}{s^2_i} + \ln(2\pi s^2_i)\right],
\end{equation}
where $\ln$ = $\log_e$, $s^2_i = \sigma^2_i + \delta^2$, where $\sigma_i$ and $\delta$ are the measurement error on the observed flux $F^{\rm obs}_{X,i}$ and the global intrinsic dispersion respectively and $F^{\rm th}_{X,i}(p)$ is the predicted flux at redshift $z_i$. The QSO data cannot constrain $H_0$ because in the $L_X-L_{UV}$ relation $\beta$ and $H_0$ have a degeneracy. In this paper we use $H_0$ as a free parameter in order to allow us to determine the complete allowed region of $\beta$.

The BAO and $H(z)$ data sets from \cite{alam} and \cite{du} are each correlated and the likelihood function for these is
\begin{equation}
\label{eq:chi2}
    \ln({\rm LF}) = -\frac{1}{2} [A_{\rm obs}(z_i) - A_{\rm th}(z_i, p)]^T \textbf{C}^{-1} [A_{\rm obs}(z_i) - A_{\rm th}(z_i, p)],
\end{equation}
where $A_{\rm obs}(z_i)$ and $A_{\rm th}(z_i, p)$ are the measured and theoretically predicted quantities respectively. For the BAO data from \cite{alam}, the covariance matrix \textbf{C} is given in eq. (19) of \cite{Khadka2020a} and the covarience matrix for the \cite{du} data is
\[
\textbf{C}=\left(
\begin{matrix}
    1.3225 & -0.1009 \\
    -0.1009 & 0.0380
\end{matrix}\right).
\]
The remaining three BAO measurements from \cite{carter2018}, \cite{DESb1}, and \cite{ata2018}, and the $H(z)$ measurements, are uncorrelated and the corresponding likelihood function is
\begin{equation}
\label{eq:chi2}
    \ln({\rm LF}) = -\frac{1}{2}\sum^{N}_{i = 1} \frac{[A_{\rm obs}(z_i) - A_{\rm th}(z_i, p)]^2}{\sigma^2_i}.
\end{equation}

To determine joint BAO and $H(z)$ constraints, we multiply the likelihoods to get a joint data likelihood. The same procedure is used to determine the joint QSO, BAO, and $H(z)$ data constraints. In our previous analyses \citep{Khadka2020a,Khadka2020b,Khadka2020c,Caokb}, in the case of the BAO data analyses, we assumed values of $\Omega_bh^2$ for the six different cosmological models from \cite{Park2018a,Park2018b,Park2018d} that were determined using CMB anisotropy data. In this paper we do not use these values of $\Omega_bh^2$ from the CMB determination, rather we let $\Omega_bh^2$ be a free parameter to be determined from the data we use in this paper.

The likelihood analyses are done using the Markov chain Monte Carlo (MCMC) method implemented in the MontePython code \citep{Brinckmann2019}. Convergence of MCMC chains for each parameter is confirmed using the Gelman-Rubin criterion $(R-1 < 0.05)$. We use a flat prior for each free parameter, non-zero over the ranges $0 \leq \Omega_bh^2 \leq 1$, $0 \leq \Omega_ch^2 \leq 1$, $0 \leq \om \leq 1$, $-2 \leq \Omega_k \leq 1$, $-5 \leq \omega_X \leq 0.33$, $0 \leq \alpha \leq 10$ , $0 \leq \delta \leq 10$, $0 \leq \beta \leq 11$, and $-5 \leq \gamma \leq 5$.\footnote{These prior ranges are sufficiently large to give stable constraints on each parameter. Constraints obtained by using QSO data are especially sensitive to the $\Omega_{k0}$ prior range. We have performed analyses of QSO data with a smaller prior range on $\Omega_{k0}$ (these results are not included in the paper) and found that in many cases this changes the constraints obtained.} For model comparison, we compute the $AIC$ and the $BIC$ values,
\begin{equation}
\label{eq:AIC}
    AIC = -2\ln(LF_{\rm max}) + 2d ,
\end{equation}
\begin{equation}
\label{eq:AIC}
    BIC = -2\ln(LF_{\rm max}) + d\ln{N},
\end{equation}
where $d$ is the number of free parameters and $N$ is the number of data points. We define the degree of freedom $dof = N - d$.

\clearpage
\onecolumn
\begin{landscape}
\addtolength{\tabcolsep}{-1.5pt}
\begin{longtable}{ccccccccccccccccc}
\caption{Unmarginalized one-dimensional best-fit parameters for all data sets.}
\label{tab:narayan}\\
\hline
Model & Data set & $\Omega_{b}h^2$ & $\Omega_{c}h^2$ & $\om$ & $\ol$ & $\ok$ & $\omega_{X}$ & $\alpha$ &$H_0$\footnotesize{$^a$}& $\delta$ & $\gamma$ & $\beta$& $\rm dof$ &-2$\ln L_{\rm max}$ & $AIC$ & $BIC$ \\
\hline
\endfirsthead
\hline
Model & Data set & $\Omega_{b}h^2$ & $\Omega_{c}h^2$ & $\om$ & $\ol$ & $\ok$ & $\omega_{X}$ & $\alpha$ &$H_0$\footnotesize{$^a$}& $\delta$ & $\gamma$ & $\beta$& $\rm dof$ &-2$\ln L_{\rm max}$ &$AIC$ & $BIC$\\
\hline
\endhead
\hline
Flat \lcdm\ & QSO$-z < 1.479$ & - & - & 0.800 & - & - & - & - &---& 0.238 & 0.584 & 8.695& 1014& 3.22 & 13.22 & 37.85\\
&QSO$-z < 1.75$ & - & - & 0.992 & - & - & - & - &---& 0.235 & 0.591 & 8.418& 1308 & $-$20.44 & $-$10.44 & 15.46\\
&QSO$-z < 2$ & - & - & 0.999 & - & - & - & - &---& 0.233 & 0.591  & 8.355 & 1529 & $-$44.84 & $-$34.84 & $-$8.16\\
&QSO$-z < 2.25$ & - & - & 0.999 & - & - & - & - &---& 0.232 & 0.594 & 8.345 & 1675 & $-$57.64 & $-$47.64 & $-$20.51\\
&QSO$-z > 1.479$ & - & - & 0.999 & - & - & - & - &---& 0.205 & 0.587 & 8.577 & 1014 & $-$154.36 & $-$144.36 & $-$119.73\\
&QSO & - & - & 0.999 & - & - & - & - &50.698& 0.226 & 0.620 & 7.701& 2033& $-$93.00  & $-$83.00 & $-$54.90 \\
&  BAO + $H(z)$& 0.024 & 0.119 & 0.298 & - & - & - & - &69.119&-&-&-& 39 & 23.66&29.66&34.87\\
& QSO-z < 1.479 + BAO + $H(z)$& 0.024 & 0.119 & 0.298 & - & - & - & - &69.213&0.239&0.599&8.300&1055&29.34&41.34&71.14\\
& QSO-z < 1.75 + BAO + $H(z)$& 0.023 & 0.121 & 0.304 & - & - & - & - &68.952&0.236&0.610&7.983& 1349& 10.10&22.10 & 53.37\\
\hline
Non-flat \lcdm\ & QSO$-z < 1.479$ & - & - & 0.768 & 1.350 & - & - &-&62.013& 0.238 & 0.580 & 8.893&1013&1.32&13.32&42.87 \\
& QSO$-z < 1.75$ & - & - & 0.701 & 1.525 & - & - &-&61.900& 0.235 & 0.578 & 8.960&1307&$-$27.48&$-$15.48&15.60\\
& QSO$-z < 2$ & - & - & 0.741 & 1.586 & - & - &-&76.901& 0.231 & 0.566 & 9.248&1528&$-$63.32&$-$51.32&$-$19.31\\
& QSO$-z < 2.25$ & - & - & 0.669 & 1.628 & - & - &-&---& 0.229 & 0.557 & 9.595 & 1674 & $-$93.00&$-$81.00& $-$48.44\\
&QSO$-z > 1.479$ & - & - &0.994 & 1.694 & - & - &-&78.805& 0.200 & 0.537 & 10.08 & 1013& $-$204.95& $-$191.95& $-$163.39 \\
& QSO & - & - & 0.900 & 1.625 & - & - &-&87.012& 0.222 & 0.558 & 9.421&2032&$-$174.44&$-$162.44&$-$128.72\\
& BAO + $H(z)$& 0.025 & 0.114 & 0.294 & 0.675 & - & - & - &68.701&-&-&-&38&23.60&31.60&38.55\\
&QSO$-z < 1.479$ + BAO + $H(z)$& 0.025 & 0.116& 0.296 & 0.692 & 0.012 & - & - &69.019&0.239&0.594&8.437&1054&29.40&43.40&78.17\\
&QSO$-z < 1.75$ + BAO + $H(z)$& 0.022 & 0.127 & 0.308 & 0.737 & $-$0.045 & - & - &69.563&0.236&0.607&8.058&1348&10.10&24.10&60.58\\
\hline
Flat XCDM & QSO$-z < 1.479$ & - & - & 0.602 & - & - & $-$4.959 &-&82.448& 0.238 & 0.588 & 8.553&1013&2.42&14.42&43.98\\
& QSO$-z < 1.75$ & - & - & 0.123 & - & - & 0.144 &-&---& 0.236 &0.586 & 8.699 & 1307&$-$21.56&$-$9.56&21.52\\
& QSO$-z < 2$ & - & - & 0.036 & - & - & 0.146 &-&48.621& 0.232 & 0.588 & 8.649&1528&$-$49.08&$-$37.08&$-$5.07\\
& QSO$-z < 2.25$ & - & - & 0.062 & - & - & 0.146 &-&56.166& 0.230 & 0.586 &8.654&1674&$-$65.24&$-$53.24&$-$20.68\\
&QSO$-z > 1.479$ & - & - & 0.321 & - & - & 0.145 &-&53.218& 0.205 & 0.581 & 8.883&1013&$-$163.26&$-$151.26&$-$121.70\\
& QSO & - & - & 0.021 & - & - & 0.149 &-&48.434& 0.225 & 0.610 & 7.982&2032&$-$109.72&$-$97.72&$-$64.00\\
&BAO + $H(z)$ & 0.031 & 0.088 & 0.280 & - & - & $-$0.691 & - &65.036& - & - & -&38&19.66&27.66&34.61\\
&QSO$-z < 1.479$ + BAO + $H(z)$ & 0.031 & 0.084 & 0.278 & - & - & $-$0.671 & - &64.398& 0.238 & 0.597 & 8.369&1054&24.98&38.98&73.75\\
&QSO$-z < 1.75$ + BAO + $H(z)$ & 0.033 & 0.081& 0.274 & - & - & $-$0.645 & - &64.394& 0.236 & 0.605 & 8.115&1348&5.68&19.68&56.16\\
\hline
Non-flat XCDM & QSO$-z < 1.479$ & - & - & 0.832 & - & $-$0.527 & $-$4.590 &-&---& 0.238 & 0.580 & 8.887&1012&1.18&15.18&49.67\\
& QSO$-z < 1.75$ & - & - & 0.164 & - & $-$1.294 & $-$0.590 &-&67.193& 0.235 & 0.572 & 9.103&1306&$-$29.38&$-$15.38&20.88\\
& QSO$-z < 2$ & - & - & 0.393 & - & $-$1.499 & $-$0.607 &-&50.567& 0.231 & 0.564 & 9.430&1527&$-$63.96&$-$49.96&$-$12.61\\
& QSO$-z < 2.25$ & - & - & 0.294 & - & $-$1.689 & $-$0.519 &-&---& 0.228 & 0.557 & 9.493&1673&$-$94.38&$-$80.38&$-$42.39\\
&QSO$-z > 1.479$ & - & - & 0.904 & - & $-$1.057 & $-$4.952 &-&63.681& 0.200 & 0.534 & 10.329 & 1012&$-$206.66&$-$192.66&$-$158.17\\
& QSO & - & - & 0.887 & - & $-$1.533 & $-$0.974 &-&51.117& 0.222 & 0.557 & 9.653& 2031& $-$174.4&$-$160.40&$-$121.06\\
& BAO + $H(z)$ & 0.030 & 0.094 & 0.291 & - & $-$0.147 & $-$0.641 & - &65.204& - & - & -&37&18.34&28.34&37.03\\
& QSO$-z < 1.479$ + BAO + $H(z)$ & 0.029 & 0.097 & 0.294 & - & $-$0.179 & $-$0.637 & - &65.731& 0.239 & 0.593 & 8.457&1053&23.10&39.10&78.84\\
& QSO$-z < 1.75$ + BAO + $H(z)$ & 0.029 & 0.097 & 0.298 & - & $-$0.217 & $-$0.614 & - &65.154& 0.236 & 0.607 & 8.054&1347&2.44&18.44&60.13\\
	\hline
Flat $\phi$CDM & QSO$-z < 1.479$ & - & - & 0.772 & - & - & - & 0.003 &52.150& 0.238 & 0.585 & 8.743&1013&3.22&15.22&44.78\\
& QSO$-z < 1.75$ & - & - & 0.999 & - & - & - & 3.150 &84.273& 0.235 & 0.591 & 8.373& 1307&$-$20.44&$-$8.44&22.64\\
& QSO$-z < 2$ & - & - & 0.998 & - & - & - & 4.667 &72.124& 0.233 & 0.591 & 8.433&1528&-44.84&-32.84&-0.83\\
& QSO$-z < 2.25$ & - & - & 0.999 & - & - & - & 6.770 &90.951& 0.231 & 0.594 & 8.280&1674&$-$57.64&$-$45.64&$-$13.08\\
& QSO$-z > 1.479$ & - & - & 0.999 & - & - & - & 3.794 &76.551& 0.206 & 0.586 & 8.632&1013&$-$154.36&$-$142.36&$-$112.80\\
& QSO & - & - & 0.999 & - & - & - & 9.929 &---& 0.226 & 0.619 & 7.671&2032&$-$93.02&$-$81.02&$-$47.30\\
& BAO + $H(z)$ & 0.033 & 0.080 & 0.265 & - & - & - & 1.445 &65.272& - & - & -&38&19.56&27.56&34.51\\
&QSO$-z < 1.479$ + BAO + $H(z)$ & 0.034 & 0.078 & 0.263 & - & - & - & 1.548 &65.542& 0.239 & 0.593 & 8.481&1054&24.82&38.82&73.59\\
&QSO$-z < 1.75$ + BAO + $H(z)$ & 0.036 & 0.069 & 0.253 & - & - & - & 1.978 &64.323& 0.237 & 0.607 & 8.053&1348&5.48&19.48&55.96\\
\hline
Non-flat $\phi$CDM & QSO$-z < 1.479$ & - & - & 0.909 & - & $-$0.833 & - & 0.008 &63.825& 0.238 & 0.583 & 8.736&1012&1.86&15.86&50.35\\
& QSO$-z < 1.75$ & - & - & 0.924 & - & $-$0.923 & - & 0.063 &87.365& 0.235 & 0.585 & 8.557&1306&$-$25.08&$-$11.08&25.18\\
& QSO$-z < 2$ & - & - & 0.998 & - & $-$0.992 & - & 0.036 &82.560& 0.231 & 0.585 & 8.592&1527&$-$55.79&$-$41.79&$-$4.44\\
& QSO$-z < 2.25$ & - & - & 0.992 & - & $-$0.988 & - & 0.046 &82.468& 0.230 & 0.578 & 8.814&1673&$-$76.60&$-$62.60&$-$24.61\\
& QSO$-z > 1.479$ & - & - & 0.995 & - & $-$0.991 & - & 0.017 &94.160& 0.203 & 0.562 & 9.272&1012&$-$181.16&$-$167.16&$-$132.67\\
& QSO & - & - & 0.997 & - & $-$0.995 & - & 0.012 &94.478& 0.224 & 0.591 & 8.366&2031&$-$137.77&$-$123.77&$-$84.43\\
& BAO + $H(z)$ & 0.035 & 0.078 & 0.261 & - & $-$0.155 & - & 2.042 &65.720& - & - & -&37&18.16&28.16&36.85\\
&QSO$-z < 1.479$ + BAO + $H(z)$ & 0.034 & 0.084 & 0.268 & - & $-$0.195 & - & 2.024 &66.289& 0.239 & 0.592 & 8.483&1053&22.84&38.84&78.58\\
&QSO$-z < 1.479$ + BAO + $H(z)$ & 0.035 & 0.077 & 0.260 & - & $-$0.246 & - & 2.404 &65.374& 0.235 & 0.603 & 8.173&1347&2.12&18.12&59.81\\
\hline
\footnotesize{$^a$  ${\rm km}\hspace{1mm}{\rm s}^{-1}{\rm Mpc}^{-1}$}\\
\end{longtable}
\end{landscape}
\clearpage
\twocolumn

\clearpage
\onecolumn
\begin{landscape}
\addtolength{\tabcolsep}{-5pt}
\begin{longtable}{ccccccccccccc}
\caption{Marginalized one-dimensional best-fit parameters with 1$\sigma$ confidence intervals for all data sets. A 2$\sigma$ limit is given when only an upper or lower limit exists.}
\label{tab:narayan1}\\
\hline
Model & Data set\hspace{5mm} & $\Omega_{b}h^2$ & $\Omega_{c}h^2$ & $\om$ & $\ol$ & $\ok$ & $\omega_{X}$ & $\alpha$ &$H_0$\footnotesize{$^a$}& $\delta$ & $\gamma$ & $\beta$ \\
\hline
\endfirsthead
\hline
Model & Data set\hspace{5mm} & $\Omega_{b}h^2$ & $\Omega_{c}h^2$ & $\om$ & $\ol$ & $\ok$ & $\omega_{X}$ & $\alpha$ &$H_0$\footnotesize{$^a$}& $\delta$ & $\gamma$ & $\beta$ \\
\hline
\endhead
Flat \lcdm\ & QSO$-z < 1.479$ & - & - & $0.670^{+0.300}_{-0.130}$ & $0.330^{+0.130}_{-0.300}$ & - & - & - &---& $0.239^{+0.006}_{-0.006}$ & $0.588^{-0.018}_{-0.018}$ & $8.570^{+0.530}_{-0.530}$\\
&QSO$-z < 1.75$ & - & - & > 0.466 & < 0.534 & - & - & - &---& $0.235^{+0.005}_{-0.005}$ & $0.595^{-0.014}_{-0.014}$ & $8.330^{+0.430}_{-0.430}$\\
&QSO$-z < 2$ & - & - & > 0.620 & < 0.380 & - & - & - &---& $0.233^{+0.005}_{-0.005}$ & $0.593^{-0.013}_{-0.013}$ & $8.400^{+0.380}_{-0.380}$\\
&QSO$-z < 2.25$ & - & - & > 0.705 & < 0.295 & - & - & - &---& $0.232^{+0.004}_{-0.004}$ & $0.595^{-0.012}_{-0.012}$ & $8.330^{+0.350}_{-0.350}$\\
&QSO$-z > 1.479$ & - & - & $> 0.642$ & $< 0.358$ & - & - & - &---& $0.206^{+0.005}_{-0.005}$ & $0.587^{-0.014}_{-0.014}$ & $8.650^{+0.420}_{-0.420}$\\
&QSO & - & - & $> 0.793$ & - & - & - & - &---& $0.227^{+0.004}_{-0.004}$ & $0.620^{-0.009}_{-0.009}$ & $7.580^{+0.290}_{-0.290}$\\
&  BAO + $H(z)$& $0.024^{+0.003}_{-0.003}$ & $0.119^{+0.008}_{-0.008}$ & $0.299^{+0.015}_{-0.017}$ & - & - & - & - &$69.300^{+1.800}_{-1.800}$&-&-&-\\
& QSO-z < 1.479 + BAO + $H(z)$& $0.024^{+0.003}_{-0.003}$ & $0.120^{+0.008}_{-0.008}$ & $0.301^{+0.015}_{-0.017}$ & - & - & - & - &$69.200^{+1.800}_{-1.800}$&$0.239^{+0.006}_{-0.006}$&$0.597^{+0.016}_{-0.016}$&$8.35^{+0.490}_{-0.490}$\\
& QSO-z < 1.75 + BAO + $H(z)$& $0.024^{+0.003}_{-0.003}$ & $0.121^{+0.008}_{-0.008}$ & $0.303^{+0.015}_{-0.017}$ & - & - & - & - &$69.100^{+1.800}_{-1.800}$&$0.236^{+0.005}_{-0.005}$&$0.609^{+0.013}_{-0.013}$&$7.990^{+0.410}_{-0.410}$\\
\hline
Non-flat \lcdm\ & QSO$-z < 1.479$ & - & - & $> 0.366$ & $0.810^{+0.810}_{-0.340}$ & $-0.510^{+0.430}_{-0.830}     $ & - &-&---& $0.239^{+0.006}_{-0.006}$ & $0.584^{+0.017}_{-0.017}$ & $8.710^{+0.530}_{-0.530}$\\
& QSO$-z < 1.75$ & - & - & $> 0.428$ & $1.300^{+0.350}_{-0.099}$ & $-1.040^{+0.210}_{-0.350}$ & - &-&---& $0.235^{+0.005}_{-0.005}$ & $0.582^{+0.016}_{-0.016}$ & $8.770^{+0.480}_{-0.480}$\\
& QSO$-z < 2$ & - & - & $> 0.464$ & $1.540^{+0.130}_{-0.077}$ & $-1.183^{+0.096}_{-0.350}$ & - &-&---& $0.231^{+0.004}_{-0.004}$ & $0.567^{+0.014}_{-0.014}$ & $9.240^{+0.430}_{-0.430}$\\
& QSO$-z < 2.25$ & - & - & $> 0.458$ & $1.610^{+0.081}_{-0.069}$ & $-1.350^{+0.160}_{-0.240}     $ & - &-&---& $0.229^{+0.004}_{-0.004}$ & $0.560^{+0.013}_{-0.013}$ & $9.450^{+0.400}_{-0.400}$\\
&QSO$-z > 1.479$ & - & - & $> 0.475$ & $1.594^{+0.095}_{-0.081}$ & $-1.370^{+0.140}_{-0.290}     $ & - &-&---& $0.201^{+0.005}_{-0.005}$ & $0.540^{+0.013}_{-0.015}$ & $10.060^{+0.460}_{-0.400}$\\
& QSO & - & - & $> 0.595$ & $1.595^{+0.064}_{-0.048}$ & $-1.410^{+0.110}_{-0.200}     $ & - &-&---& $0.222^{+0.004}_{-0.004}$ & $0.560^{+0.012}_{-0.012}$ & $9.460^{+0.370}_{-0.370}$\\
& BAO + $H(z)$& $0.025^{+0.004}_{-0.004}$ & $0.113^{+0.019}_{-0.019}$ & $0.292^{+0.023}_{-0.023}$ & $0.667^{+0.093}_{+0.081}$ & $-0.014^{+0.075}_{-0.075}$ & - & - &$68.700^{+2.300}_{-2.300}$&-&-&-\\
&QSO$-z < 1.479$ + BAO + $H(z)$& $0.025^{+0.005}_{-0.005}$ & $0.117^{+0.019}_{-0.019}$ & $0.297^{+0.024}_{-0.024}$ & $0.680^{+0.092}_{+0.079}$ & $0.023^{+0.094}_{-0.011}$ & - & - &$68.900^{+2.200}_{-2.200}$&$0.239^{+0.006}_{-0.006}$&$0.597^{+0.016}_{-0.016}$&$8.350^{+0.480}_{-0.480}$\\
&QSO$-z < 1.75$ + BAO + $H(z)$& $0.023^{+0.004}_{-0.004}$ & $0.123^{+0.018}_{-0.018}$ & $0.305^{+0.023}_{-0.023}$ & $0.705^{+0.090}_{+0.078}$ & $-0.010^{+0.092}_{-0.110}$ & - & - &$68.900^{+2.200}_{-2.200}$&$0.239^{+0.006}_{-0.006}$&$0.597^{+0.016}_{-0.016}$&$8.350^{+0.480}_{-0.480}$\\
\hline
Flat XCDM & QSO$-z < 1.479$ & - & - & $0.650^{+0.270}_{-0.160}$ & - & - & < $-$0.137 &-&---& $0.239^{+0.006}_{-0.006}$ & $0.588^{+0.018}_{-0.018}$ & $8.580^{+0.590}_{-0.520}$\\
& QSO$-z < 1.75$ & - & - & $> 0.361$ & - & - & < $-$0.0126 &-&---& $0.236^{+0.005}_{-0.005}$ & $0.595^{+0.014}_{-0.014}$ & $8.350^{+0.430}_{-0.430}$\\
& QSO$-z < 2$ & - & - & --- & - & - & < 0.115 &-&---& $0.233^{+0.005}_{-0.005}$ & $0.593^{+0.013}_{-0.013}$ & $8.430^{+0.3900}_{-0.3900}$\\
& QSO$-z < 2.25$ & - & - & --- & - & - & < 0.137 &-&---& $0.232^{+0.004}_{-0.004}$ & $0.594^{+0.012}_{-0.012}$ & $8.390^{+0.370}_{-0.370}$\\
&QSO$-z > 1.479$ & - & - & --- & - & - & < 0.140 &-&---& $0.206^{+0.006}_{-0.006}$ & $0.586^{+0.015}_{-0.015}$ & $8.690^{+0.450}_{-0.450}$\\
& QSO & - & - & < 0.247 & - & - & $0.115^{+0.036}_{-0.014}$ &-&---& $0.226^{+0.004}_{-0.004}$ & $0.612^{+0.009}_{-0.009}$ & $7.820^{+0.290}_{-0.290}$\\
&BAO + $H(z)$ & $0.030^{+0.005}_{-0.005}$ & $0.093^{+0.019}_{-0.017}$ & $0.282^{+0.021}_{-0.021}$ & - & - & $-0.744^{+0.140}_{-0.097}$ & - &$65.800^{+2.200}_{-2.500}$& - & - & -\\
&QSO$-z < 1.479$ + BAO + $H(z)$ & $0.030^{+0.005}_{-0.005}$ & $0.092^{+0.019}_{-0.017}$ & $0.282^{+0.023}_{-0.021}$ & - & - & $-0.7310^{+0.150}_{-0.095}$ & - &$65.500^{+2.200}_{-2.500}$& $0.239^{+0.006}_{-0.006}$ & $0.595^{+0.016}_{-0.016}$ & $8.420^{+0.490}_{-0.490}$\\
&QSO$-z < 1.75$ + BAO + $H(z)$ & $0.030^{+0.005}_{-0.005}$ & $0.092^{+0.020}_{-0.017}$ & $0.284^{+0.024}_{-0.021}$ & - & - & $-0.7260^{+0.150}_{-0.096}$ & - &$65.400^{+2.200}_{-2.500}$& $0.236^{+0.005}_{-0.005}$ & $0.608^{+0.013}_{-0.013}$ & $8.040^{+0.410}_{-0.410}$\\
\hline
Non-flat XCDM & QSO$-z < 1.479$ & - & - & > 0.285 & - & $-0.230^{+0.520}_{-0.450}$ & $-2.300^{+2.000}_{-1.500}$ &-&---& $0.239^{+0.005}_{-0.006}$ & $0.587^{+0.019}_{-0.016}$ & $8.610^{+0.520}_{-0.520}$\\
& QSO$-z < 1.75$ & - & - & > 0.306 & - & $-0.740^{+0.560}_{-0.440}$ & $-2.100^{+2.000}_{-1.000}$ &-&---& $0.235^{+0.005}_{-0.005}$ & $0.585^{+0.016}_{-0.016}$ & $8.700^{+0.490}_{-0.490}$\\
& QSO$-z < 2$ & - & - & > 0.268 & - & $-1.180^{+0.500}_{-0.400}$ & $-1.740^{+1.700}_{-0.590}$ &-&---& $0.232^{+0.005}_{-0.005}$ & $0.569^{+0.015}_{-0.015}$ & $9.180^{+0.450}_{-0.450}$\\
& QSO$-z < 2.25$ & - & - & > 0.204 & - & $-1.390^{+0.290}_{-0.500}$ & $-1.250^{+1.000}_{-0.260}$ &-&---& $0.229^{+0.004}_{-0.004}$ & $0.560^{+0.014}_{-0.014}$ & $9.430^{+0.420}_{-0.420}$\\
&QSO$-z > 1.479$ & - & - & >0.333 & - & $-0.940^{+0.290}_{-0.290}$ & $-3.000^{+1.000}_{-1.600}$ &-&---& $0.201^{+0.005}_{-0.005}$ & $0.541^{+0.013}_{-0.016}$ & $10.130^{+0.510}_{-0.420}$\\
& QSO & - & - & > 0.579 & - & $-1.350^{+0.350}_{-0.290}$ & $-1.360^{+0.790}_{-0.250}$ &-&---& $0.222^{+0.004}_{-0.004}$ & $0.560^{+0.012}_{-0.012}$ & $9.460^{+0.360}_{-0.360}$\\
& BAO + $H(z)$ & $0.029^{+0.005}_{-0.005}$ & $0.099^{+0.021}_{-0.021}$ & $0.293^{+0.027}_{-0.027}$ & - & $-0.120^{+0.130}_{-0.130}$ & $-0.693^{+0.130}_{-0.077}$ & - &$65.900^{+2.400}_{-2.400}$& - & - & -\\
& QSO$-z < 1.479$ + BAO + $H(z)$ & $0.029^{+0.005}_{-0.006}$ & $0.099^{+0.021}_{-0.021}$ & $0.295^{+0.028}_{-0.028}$ & - & $-0.150^{+0.130}_{-0.130}$ & $-0.675^{+0.130}_{-0.075}$ & - &$65.800^{+2.300}_{-2.300}$& $0.239^{+0.006}_{-0.006}$ & $0.593^{+0.016}_{-0.016}$ & $8.460^{+0.490}_{-0.490}$\\
& QSO$-z < 1.75$ + BAO + $H(z)$ & $0.029^{+0.005}_{-0.006}$ & $0.102^{+0.022}_{-0.022}$ & $0.300^{+0.028}_{-0.028}$ & - & $-0.210^{+0.130}_{-0.130}$ & $-0.653^{+0.120}_{-0.069}$ & - &$65.800^{+2.300}_{-2.300}$& $0.236^{+0.005}_{-0.005}$ & $0.604^{+0.014}_{-0.014}$ & $8.130^{+0.420}_{-0.420}$\\
\hline
Flat $\phi$CDM & QSO$-z < 1.479$ & - & - & $0.600^{+0.340}_{-0.170}$ & - & - & - & $5.200^{+3.800}_{-2.500}$ &---& $0.239^{+0.006}_{-0.006}$ & $0.586^{-0.017}_{-0.017}$ & $8.610^{+0.510}_{-0.510}$\\
& QSO$-z < 1.75$ & - & - & >0.342 & - & - & - & $5.300^{+4.200}_{-1.900}$ &---& $0.236^{+0.005}_{-0.005}$ & $0.595^{-0.014}_{-0.014}$ & $8.340^{+0.430}_{-0.430}$\\
& QSO$-z < 2$ & - & - & >0.511 & - & - & - & --- &---& $0.233^{+0.004}_{-0.004}$ & $0.593^{-0.012}_{-0.012}$ & $8.410^{+0.380}_{-0.380}$\\
& QSO$-z < 2.25$ & - & - & >0.627 & - & - & - & --- &---& $0.232^{+0.004}_{-0.004}$ & $0.595^{-0.012}_{-0.012}$ & $8.340^{+0.350}_{-0.350}$\\
& QSO$-z > 1.479$ & - & - & > 0.614 & - & - & - & --- &---& $0.206^{+0.005}_{-0.005}$ & $0.587^{-0.014}_{-0.014}$ & $8.640^{+0.420}_{-0.420}$\\
& QSO & - & - & $> 0.756$ & - & - & - & --- &---& $0.227^{+0.004}_{-0.004}$ & $0.620^{-0.009}_{-0.009}$ & $7.590^{+0.290}_{-0.290}$\\
& BAO + $H(z)$ & $0.032^{+0.006}_{-0.003}$ & $0.081^{+0.017}_{-0.017}$ & $0.266^{+0.023}_{-0.023}$ & - & - & - & $1.530^{+0.620}_{-0.850}$ &$65.100^{+2.100}_{-2.100}$& - & - & -\\
&QSO$-z < 1.479$ + BAO + $H(z)$ & $0.033^{+0.007}_{-0.003}$ & $0.080^{+0.017}_{-0.019}$ & $0.266^{+0.024}_{-0.024}$ & - & - & - & $1.590^{+0.650}_{-0.880}$ &$65.000^{+2.100}_{-2.100}$& $0.239^{+0.006}_{-0.006}$ & $0.595^{+0.016}_{-0.016}$ & $8.430^{+0.490}_{-0.490}$\\
&QSO$-z < 1.75$ + BAO + $H(z)$ & $0.033^{+0.006}_{-0.003}$ & $0.080^{+0.017}_{-0.020}$ & $0.267^{+0.024}_{-0.024}$ & - & - & - & $1.620^{+0.670}_{-0.880}$ &$64.800^{+2.100}_{-2.100}$& $0.236^{+0.005}_{-0.005}$ & $0.607^{+0.014}_{-0.014}$ & $8.050^{+0.410}_{-0.410}$\\
\hline
Non-flat $\phi$CDM & QSO$-z < 1.479$ & - & - & $0.610^{+0.320}_{-0.170}$ & - & $-0.110^{+0.330}_{-0.330}$ & - & --- &---& $0.239^{+0.006}_{-0.006}$ & $0.585^{-0.017}_{-0.017}$ & $8.630^{+0.510}_{-0.510}$\\
& QSO$-z < 1.75$ & - & - & > 0.418 & - & $-0.340^{+0.300}_{-0.300}$ & - & --- &---& $0.236^{+0.005}_{-0.005}$ & $0.591^{-0.015}_{-0.015}$ & $8.460^{+0.440}_{-0.440}$\\
& QSO$-z < 2$ & - & - & > 0.613 & - & $-0.610^{+0.120}_{-0.320}$ & - & --- &---& $0.233^{+0.005}_{-0.005}$ & $0.585^{-0.013}_{-0.013}$ & $8.630^{+0.390}_{-0.390}$\\
& QSO$-z < 2.25$ & - & - & > 0.731 & - & $-0.770^{+0.078}_{-0.190}$ & - & --- &---& $0.231^{+0.004}_{-0.004}$ & $0.584^{-0.012}_{-0.012}$ & $8.650^{+0.370}_{-0.370}$\\
& QSO$-z > 1.479$ & - & - & > 0.745 & - & $-0.828^{+0.062}_{-0.140}$ & - & < 8.390 &---& $0.204^{+0.005}_{-0.005}$ & $0.572^{-0.014}_{-0.014}$ & $9.090^{+0.450}_{-0.450}$\\
& QSO & - & - & $> 0.865$ & - & $-0.913^{+0.110}_{-0.092}$ & - & <4.590 &---& $0.224^{+0.004}_{-0.004}$ & $0.599^{-0.010}_{-0.010}$ & $8.240^{+0.310}_{-0.310}$\\
& BAO + $H(z)$ & $0.032^{+0.006}_{-0.004}$ & $0.085^{+0.017}_{-0.021}$ & $0.271^{+0.024}_{-0.028}$ & - & $-0.080^{+0.100}_{-0.100}$ & - & $1.660^{+0.670}_{-0.830}$ &$65.500^{+2.500}_{-2.500}$& - & - & -\\
&QSO$-z < 1.479$ + BAO + $H(z)$ & $0.032^{+0.007}_{-0.003}$ & $0.085^{+0.018}_{-0.023}$ & $0.272^{+0.024}_{-0.029}$ & - & $-0.104^{+0.084}_{-0.130}$ & - & $1.760^{+0.700}_{-0.820}$ &$65.500^{+2.200}_{-2.200}$& $0.239^{+0.006}_{-0.006}$ & $0.594^{+0.016}_{-0.016}$ & $8.450^{+0.490}_{-0.490}$\\
&QSO$-z < 1.479$ + BAO + $H(z)$ & $0.032^{+0.007}_{-0.004}$ & $0.088^{+0.019}_{-0.023}$ & $0.276^{+0.025}_{-0.030}$ & - & $-0.138^{+0.074}_{-0.120}$ & - & $1.820^{+0.680}_{-0.810}$ &$65.600^{+2.200}_{-2.200}$& $0.236^{+0.005}_{-0.005}$ & $0.605^{+0.014}_{-0.014}$ & $8.110^{+0.410}_{-0.410}$\\
\hline
\footnotesize{$^a$  ${\rm km}\hspace{1mm}{\rm s}^{-1}{\rm Mpc}^{-1}$}\\
\end{longtable}
\end{landscape}
\clearpage
\twocolumn

\begin{table}
	\centering
	\small\addtolength{\tabcolsep}{-2.5pt}
	\small
	\caption{$L_X-L_{UV}$ relation parameters (and $\delta$) differences between different models for the full QSO data.}
	\label{tab:BFP}
	\begin{threeparttable}
	\begin{tabular}{lccccccccccc} 
		\hline
		Model vs model & $\Delta \delta$ & $\Delta \gamma$ & $\Delta \beta$\\
		\hline
		Flat \lcdm\ vs non-flat \lcdm\ & $0.88\sigma$ & $4.00\sigma$ & $4.00\sigma$\\
		Flat \lcdm\ vs flat XCDM & $0.18\sigma$ & $0.63\sigma$ & $0.59\sigma$\\
		Flat \lcdm\ vs non-flat XCDM & $0.88\sigma$ & $4.00\sigma$ & $4.07\sigma$\\
		Flat \lcdm\ vs flat $\phi$CDM & $0.00\sigma$ & $0.00\sigma$ & $0.02\sigma$\\
		Flat \lcdm\ vs non-flat $\phi$CDM & $0.53\sigma$ & $1.56\sigma$ & $1.55\sigma$\\
		Non-flat \lcdm\ vs flat XCDM & $0.71\sigma$ & $3.47\sigma$ & $3.49\sigma$\\
		Non-flat \lcdm\ vs non-flat XCDM & $0.00\sigma$ & $0.00\sigma$ & $0.00\sigma$\\
		Non-flat \lcdm\ vs flat $\phi$CDM & $0.88\sigma$ & $4.00\sigma$ & $3.98\sigma$\\
		Non-flat \lcdm\ vs non-flat $\phi$CDM & $0.35\sigma$ & $2.50\sigma$ & $2.53\sigma$\\
		Flat XCDM vs non-flat XCDM & $0.71\sigma$ & $3.47\sigma$ & $3.55\sigma$\\
		Flat XCDM vs flat $\phi$CDM & $0.18\sigma$ & $0.63\sigma$ & $0.56\sigma$\\
		Flat XCDM vs non-flat $\phi$CDM & $0.35\sigma$ & $0.97\sigma$ & $0.99\sigma$\\
		Non-flat XCDM vs flat $\phi$CDM & $0.88\sigma$ & $4.00\sigma$ & $4.06\sigma$\\
		Non-flat XCDM vs non-flat $\phi$CDM & $0.35\sigma$ & $2.50\sigma$ & $2.57\sigma$\\
		Flat $\phi$CDM vs non-flat $\phi$CDM & $0.53\sigma$ & $1.56\sigma$ & $1.53\sigma$\\
		\hline
	\end{tabular}
    \end{threeparttable}
\end{table}

\begin{table}
\small\addtolength{\tabcolsep}{1.5pt}
\caption{$L_X-L_{UV}$ relation parameters (and $\delta$) differences between different models and different QSO redshift data sets.}
\label{tab:BFP}
\begin{tabular}{cccc} 
\hline
Model & \hspace{8mm}$\Delta \delta$ \hspace{8mm} & \hspace{8mm}$\Delta \gamma$ \hspace{8mm} & $\Delta \beta$\\
\hline
\multicolumn{4}{c}{Between QSO$-z < 1.479$ and QSO$-z > 1.479$} \\
\hline
Flat \lcdm\ & $4.23\sigma$ & $0.04\sigma$ & $0.12\sigma$\\
Non-flat \lcdm\ & $4.87\sigma$ & $2.06\sigma$ & $2.03\sigma$ \\
Flat XCDM & $3.89\sigma$ & $0.09\sigma$ & $0.15\sigma$\\
Non-flat XCDM & $4.87\sigma$ & $2.23\sigma$ & $2.27\sigma$\\
Flat $\phi$CDM & $4.23\sigma$ & $0.05\sigma$ & $0.05\sigma$\\
Non-flat $\phi$CDM & $4.48\sigma$ & $0.59\sigma$ & $0.68\sigma$\\
\hline
\multicolumn{4}{c}{Between QSO$-z < 1.479$ and QSO} \\
\hline
Flat \lcdm\ & $1.66\sigma$ & $1.59\sigma$ & $1.64\sigma$\\
Non-flat \lcdm\ & $2.36\sigma$ & $1.15\sigma$ & $1.16\sigma$ \\
Flat XCDM & $2.23\sigma$ & $1.45\sigma$ & $1.51\sigma$\\
Non-flat XCDM & $2.08\sigma$ & $1.35\sigma$ & $1.34\sigma$\\
Flat $\phi$CDM & $1.66\sigma$ & $1.77\sigma$ & $1.74\sigma$\\
Non-flat $\phi$CDM & $2.08\sigma$ & $0.71\sigma$ & $0.65\sigma$\\
\hline
\multicolumn{4}{c}{Between QSO$-z > 1.479$ and QSO} \\
\hline
Flat \lcdm\ & $3.28\sigma$ & $1.98\sigma$ & $2.10\sigma$\\
Non-flat \lcdm\ & $3.28\sigma$ & $1.13\sigma$ & $1.10\sigma$ \\
Flat XCDM & $2.77\sigma$ & $1.49\sigma$ & $1.61\sigma$\\
Non-flat XCDM & $3.28\sigma$ & $1.07\sigma$ & $1.21\sigma$\\
Flat $\phi$CDM & $3.28\sigma$ & $1.98\sigma$ & $2.05\sigma$\\
Non-flat $\phi$CDM & $3.12\sigma$ & $1.57\sigma$ & $1.56\sigma$\\
\hline
\end{tabular}
\end{table}

\begin{table}
	\centering
	\small\addtolength{\tabcolsep}{-2.5pt}
	\small
	\caption{$L_X-L_{UV}$ relation parameters (and $\delta$) differences between different models for QSO$-z < 1.479$ data.}
	\label{tab:BFP}
	\begin{threeparttable}
	\begin{tabular}{lccccccccccc} 
		\hline
		Model vs model & $\Delta \delta$ & $\Delta \gamma$ & $\Delta \beta$\\
		\hline
		Flat \lcdm\ vs non-flat \lcdm\ & $0.00\sigma$ & $0.16\sigma$ & $0.19\sigma$\\
		Flat \lcdm\ vs flat XCDM & $0.00\sigma$ & $0.00\sigma$ & $0.01\sigma$\\
		Flat \lcdm\ vs non-flat XCDM & $0.00\sigma$ & $0.04\sigma$ & $0.05\sigma$\\
		Flat \lcdm\ vs flat $\phi$CDM & $0.00\sigma$ & $0.08\sigma$ & $0.05\sigma$\\
		Flat \lcdm\ vs non-flat $\phi$CDM & $0.00\sigma$ & $0.12\sigma$ & $0.08\sigma$\\
		Non-flat \lcdm\ vs flat XCDM & $0.00\sigma$ & $0.16\sigma$ & $0.16\sigma$\\
		Non-flat \lcdm\ vs non-flat XCDM & $0.00\sigma$ & $0.13\sigma$ & $0.13\sigma$\\
		Non-flat \lcdm\ vs flat $\phi$CDM & $0.00\sigma$ & $0.08\sigma$ & $0.05\sigma$\\
		Non-flat \lcdm\ vs non-flat $\phi$CDM & $0.00\sigma$ & $0.04\sigma$ & $0.11\sigma$\\
		Flat XCDM vs non-flat XCDM & $0.00\sigma$ & $0.04\sigma$ & $0.04\sigma$\\
		Flat XCDM vs flat $\phi$CDM & $0.00\sigma$ & $0.08\sigma$ & $0.04\sigma$\\
		Flat XCDM vs non-flat $\phi$CDM & $0.00\sigma$ & $0.12\sigma$ & $0.06\sigma$\\
		Non-flat XCDM vs flat $\phi$CDM & $0.00\sigma$ & $0.04\sigma$ & $0.00\sigma$\\
		Non-flat XCDM vs non-flat $\phi$CDM & $0.00\sigma$ & $0.08\sigma$ & $0.03\sigma$\\
		Flat $\phi$CDM vs non-flat $\phi$CDM & $0.00\sigma$ & $0.04\sigma$ & $0.03\sigma$\\
		\hline
	\end{tabular}
    \end{threeparttable}
\end{table}

\begin{table}
	\centering
	\small\addtolength{\tabcolsep}{-2.5pt}
	\small
	\caption{$L_X-L_{UV}$ relation parameters (and $\delta$) differences between different models for QSO$-z > 1.479$ data.}
	\label{tab:BFP}
	\begin{threeparttable}
	\begin{tabular}{lccccccccccc} 
		\hline
		Model vs model & $\Delta \delta$ & $\Delta \gamma$ & $\Delta \beta$\\
		\hline
		Flat \lcdm\ vs non-flat \lcdm\ & $0.71\sigma$ & $2.46\sigma$ & $2.43\sigma$\\
		Flat \lcdm\ vs flat XCDM & $0.00\sigma$ & $0.05\sigma$ & $0.06\sigma$\\
		Flat \lcdm\ vs non-flat XCDM & $0.71\sigma$ & $2.41\sigma$ & $2.49\sigma$\\
		Flat \lcdm\ vs flat $\phi$CDM & $0.00\sigma$ & $0.00\sigma$ & $0.02\sigma$\\
		Flat \lcdm\ vs non-flat $\phi$CDM & $0.28\sigma$ & $0.76\sigma$ & $0.71\sigma$\\
		Non-flat \lcdm\ vs flat XCDM & $0.64\sigma$ & $2.32\sigma$ & $2.28\sigma$\\
		Non-flat \lcdm\ vs non-flat XCDM & $0.00\sigma$ & $0.05\sigma$ & $0.11\sigma$\\
		Non-flat \lcdm\ vs flat $\phi$CDM & $0.71\sigma$ & $2.46\sigma$ & $2.45\sigma$\\
		Non-flat \lcdm\ vs non-flat $\phi$CDM & $0.42\sigma$ & $1.67\sigma$ & $1.61\sigma$\\
		Flat XCDM vs non-flat XCDM & $0.64\sigma$ & $2.27\sigma$ & $2.34\sigma$\\
		Flat XCDM vs flat $\phi$CDM & $0.00\sigma$ & $0.05\sigma$ & $0.08\sigma$\\
		Flat XCDM vs non-flat $\phi$CDM & $0.26\sigma$ & $0.68\sigma$ & $0.63\sigma$\\
		Non-flat XCDM vs flat $\phi$CDM & $0.71\sigma$ & $2.41\sigma$ & $2.51\sigma$\\
		Non-flat XCDM vs non-flat $\phi$CDM & $0.42\sigma$ & $1.62\sigma$ & $1.69\sigma$\\
		Flat $\phi$CDM vs non-flat $\phi$CDM & $0.28\sigma$ & $0.76\sigma$ & $0.73\sigma$\\
		\hline
	\end{tabular}
    \end{threeparttable}
\end{table}

\begin{table}
	\centering
	\small\addtolength{\tabcolsep}{-2.5pt}
	\small
	\caption{$L_X-L_{UV}$ relation parameters (and $\delta$) differences between different models for QSO$-z < 1.75$ data.}
	\label{tab:BFP}
	\begin{threeparttable}
	\begin{tabular}{lccccccccccc} 
		\hline
		Model vs model & $\Delta \delta$ & $\Delta \gamma$ & $\Delta \beta$\\
		\hline
		Flat \lcdm\ vs non-flat \lcdm\ & $0.00\sigma$ & $0.61\sigma$ & $0.68\sigma$\\
		Flat \lcdm\ vs flat XCDM & $0.14\sigma$ & $0.00\sigma$ & $0.03\sigma$\\
		Flat \lcdm\ vs non-flat XCDM & $0.00\sigma$ & $0.47\sigma$ & $0.57\sigma$\\
		Flat \lcdm\ vs flat $\phi$CDM & $0.14\sigma$ & $0.00\sigma$ & $0.02\sigma$\\
		Flat \lcdm\ vs non-flat $\phi$CDM & $0.14\sigma$ & $0.19\sigma$ & $0.21\sigma$\\
		Non-flat \lcdm\ vs flat XCDM & $0.14\sigma$ & $0.61\sigma$ & $0.65\sigma$\\
		Non-flat \lcdm\ vs non-flat XCDM & $0.00\sigma$ & $0.13\sigma$ & $0.10\sigma$\\
		Non-flat \lcdm\ vs flat $\phi$CDM & $0.14\sigma$ & $0.61\sigma$ & $0.67\sigma$\\
		Non-flat \lcdm\ vs non-flat $\phi$CDM & $0.14\sigma$ & $0.41\sigma$ & $0.48\sigma$\\
		Flat XCDM vs non-flat XCDM & $0.14\sigma$ & $0.47\sigma$ & $0.54\sigma$\\
		Flat XCDM vs flat $\phi$CDM & $0.00\sigma$ & $0.00\sigma$ & $0.02\sigma$\\
		Flat XCDM vs non-flat $\phi$CDM & $0.00\sigma$ & $0.19\sigma$ & $0.18\sigma$\\
		Non-flat XCDM vs flat $\phi$CDM & $0.14\sigma$ & $0.47\sigma$ & $0.55\sigma$\\
		Non-flat XCDM vs non-flat $\phi$CDM & $0.14\sigma$ & $0.27\sigma$ & $0.36\sigma$\\
		Flat $\phi$CDM vs non-flat $\phi$CDM & $0.00\sigma$ & $0.19\sigma$ & $0.20\sigma$\\
		\hline
	\end{tabular}
    \end{threeparttable}
\end{table}

\begin{table}
	\centering
	\small\addtolength{\tabcolsep}{-2.5pt}
	\small
	\caption{$L_X-L_{UV}$ relation parameters (and $\delta$) differences between different models for QSO$-z < 2$ data.}
	\label{tab:BFP}
	\begin{threeparttable}
	\begin{tabular}{lccccccccccc} 
		\hline
		Model vs model & $\Delta \delta$ & $\Delta \gamma$ & $\Delta \beta$\\
		\hline
		Flat \lcdm\ vs non-flat \lcdm\ & $0.31\sigma$ & $1.36\sigma$ & $1.46\sigma$\\
		Flat \lcdm\ vs flat XCDM & $0.00\sigma$ & $0.00\sigma$ & $0.06\sigma$\\
		Flat \lcdm\ vs non-flat XCDM & $0.14\sigma$ & $1.21\sigma$ & $1.32\sigma$\\
		Flat \lcdm\ vs flat $\phi$CDM & $0.00\sigma$ & $0.00\sigma$ & $0.02\sigma$\\
		Flat \lcdm\ vs non-flat $\phi$CDM & $0.00\sigma$ & $0.44\sigma$ & $0.42\sigma$\\
		Non-flat \lcdm\ vs flat XCDM & $0.31\sigma$ & $1.36\sigma$ & $1.40\sigma$\\
		Non-flat \lcdm\ vs non-flat XCDM & $0.15\sigma$ & $0.10\sigma$ & $0.10\sigma$\\
		Non-flat \lcdm\ vs flat $\phi$CDM & $0.35\sigma$ & $1.36\sigma$ & $1.45\sigma$\\
		Non-flat \lcdm\ vs non-flat $\phi$CDM & $0.31\sigma$ & $0.94\sigma$ & $1.05\sigma$\\
		Flat XCDM vs non-flat XCDM & $0.14\sigma$ & $1.21\sigma$ & $1.26\sigma$\\
		Flat XCDM vs flat $\phi$CDM & $0.00\sigma$ & $0.00\sigma$ & $0.04\sigma$\\
		Flat XCDM vs non-flat $\phi$CDM & $0.00\sigma$ & $0.44\sigma$ & $0.36\sigma$\\
		Non-flat XCDM vs flat $\phi$CDM & $0.16\sigma$ & $1.25\sigma$ & $1.31\sigma$\\
		Non-flat XCDM vs flat $\phi$CDM & $0.14\sigma$ & $0.81\sigma$ & $0.92\sigma$\\
		Flat $\phi$CDM vs non-flat $\phi$CDM & $0.00\sigma$ & $0.40\sigma$ & $0.40\sigma$\\
		\hline
	\end{tabular}
    \end{threeparttable}
\end{table}

\begin{table}
	\centering
	\small\addtolength{\tabcolsep}{-2.5pt}
	\small
	\caption{$L_X-L_{UV}$ relation parameters (and $\delta$) differences between different models for QSO$-z < 2.25$ data.}
	\label{tab:BFP}
	\begin{threeparttable}
	\begin{tabular}{lccccccccccc} 
		\hline
		Model vs model & $\Delta \delta$ & $\Delta \gamma$ & $\Delta \beta$\\
		\hline
		Flat \lcdm\ vs non-flat \lcdm\ & $0.53\sigma$ & $1.98\sigma$ & $2.11\sigma$\\
		Flat \lcdm\ vs flat XCDM & $0.00\sigma$ & $0.06\sigma$ & $0.12\sigma$\\
		Flat \lcdm\ vs non-flat XCDM & $0.53\sigma$ & $1.90\sigma$ & $2.01\sigma$\\
		Flat \lcdm\ vs flat $\phi$CDM & $0.00\sigma$ & $0.00\sigma$ & $0.02\sigma$\\
		Flat \lcdm\ vs non-flat $\phi$CDM & $0.00\sigma$ & $0.06\sigma$ & $0.12\sigma$\\
		Non-flat \lcdm\ vs flat XCDM & $0.53\sigma$ & $1.92\sigma$ & $1.95\sigma$\\
		Non-flat \lcdm\ vs non-flat XCDM & $0.00\sigma$ & $0.00\sigma$ & $0.03\sigma$\\
		Non-flat \lcdm\ vs flat $\phi$CDM & $0.53\sigma$ & $1.98\sigma$ & $2.09\sigma$\\
		Non-flat \lcdm\ vs flat $\phi$CDM & $0.35\sigma$ & $1.36\sigma$ & $1.47\sigma$\\
		Flat XCDM vs non-flat XCDM & $0.53\sigma$ & $1.90\sigma$ & $1.86\sigma$\\
		Flat XCDM vs flat $\phi$CDM & $0.00\sigma$ & $0.06\sigma$ & $0.10\sigma$\\
		Flat XCDM vs non-flat $\phi$CDM & $0.18\sigma$ & $0.59\sigma$ & $0.50\sigma$\\
		Non-flat XCDM vs flat $\phi$CDM & $0.53\sigma$ & $1.90\sigma$ & $1.99\sigma$\\
		Non-flat XCDM vs non-flat $\phi$CDM & $0.35\sigma$ & $1.30\sigma$ & $1.39\sigma$\\
		Flat $\phi$CDM vs non-flat $\phi$CDM & $0.18\sigma$ & $0.65\sigma$ & $0.61\sigma$\\
		\hline
	\end{tabular}
    \end{threeparttable}
\end{table}


\section{Results}
\label{sec:QSO}
\subsection{QSO data consistency tests}
\label{QSO}
Cosmological model and $L_X-L_{UV}$ relation parameter constraints from the full (2038) QSO data are listed in Tables 2 and 3 (these are in the lines labeled "QSO" in the second column of these tables) and one-dimensional likelihoods and two-dimensional constraint contours are plotted in Figs.\ 2--4. Constraints from various subsets of QSO data are also listed in these tables and plotted in Figs.\ 5--7.

In most of the cosmological models, the full QSO data favor very high values of $\Omega_{m0}$. The values range from > 0.569 to > 0.865 at the 2$\sigma$ lower limit. Surprisingly, in the flat XCDM parametrization the $\Omega_{m0}$ value is determined to be < 0.247 at the 2$\sigma$ upper limit. These values are not consistent with estimates from other well established cosmological probes \citep{chen03,Plank2018}. This was also an issue with the 2019 \citep{Risaliti2019} data compilation \citep{Khadka2020b}

Comparing the $\gamma$ and $\beta$ values listed in the last two columns of Table 3, for each of the six cosmological models, for the full QSO data set lines, we see that these are significantly model dependent. This means that the $L_X-L_{UV}$ relation of eq.\ (7) depends on the cosmological model, which means that these QSOs cannot be used to constrain cosmological parameters. This is more clearly illustrated in Table 4 which lists the differences between pairs of $\gamma$, $\beta$, and $\delta$ values determined in each of the 15 pairs of models, in terms of the quadrature sum of the two error bars in each pairs. The difference between $\delta$ values ($\Delta \delta$) ranges between (0 -- 0.88)$\sigma$ which is not statistically significant. The difference between $\gamma$ values ($\Delta \gamma$) ranges between (0 -- 4)$\sigma$. Also, the difference between $\beta$ values ($\Delta \beta$) from model to model ranges between (0 -- 4.1)$\sigma$. These differences in $\gamma$ and $\beta$ values show that for the full QSO data the determined $\gamma$ and $\beta$ values are model dependent. Consequently, the $L_X-L_{UV}$ relation cannot standardize all QSOs in the full (2038) QSO data set. We note that $\gamma$ and $\beta$ values determined from the 2015 and 2019 QSO compilations \citep{Risaliti2015,Risaliti2019} are independent of the cosmological model used in the analysis \citep{Khadka2020a,Khadka2020b}.

It is extremely curious that $\Delta \gamma$ and $\Delta \beta$ are quite small for four pairs of models, the flat $\Lambda$CDM-XCDM, $\Lambda$CDM-$\phi$CDM, and XCDM-$\phi$CDM pairs (only two of which are independent) and the non-flat $\Lambda$CDM-XCDM pair. Perhaps the first three spatially-flat pairs agreements might be understandable if the QSO standardization only worked in a spatially-flat geometry, which could explain the large differences for flat and non-flat model pairs, but this explanation does not seem to be consistent with the good agreement between non-flat $\Lambda$CDM and XCDM and the large discrepancy between non-flat $\Lambda$CDM and $\phi$CDM or between non-flat $\phi$CDM and XCDM. These patterns are not exclusive to just the full (2038) QSO data set; they hold also for most of the other QSO data subsets. 

In Table 5, we compare $\delta$, $\gamma$, and $\beta$ values from the QSO, QSO$-z < 1.479$, and QSO$-z > 1.479$ data sets. The first group contains all 2038 quasars, and other two contain the 1019 low and 1019 high redshift quasars. The QSO$-z < 1.479$ and QSO$-z > 1.479$ $\delta$, $\gamma$, and $\beta$ values are listed in Table 3 in the correspondingly labeled lines (see second column in the table). In the QSO$-z < 1.479$ and QSO$-z > 1.479$ comparison, the difference between $\delta$ values ($\Delta \delta$) ranges between (3.9 -- 4.9)$\sigma$ which is a very large difference. The difference between $\gamma$ values ($\Delta \gamma$) ranges between (0.04 -- 2.2)$\sigma$ which can be a significant difference. The difference between $\beta$ values ($\Delta \beta$) ranges between (0.05 -- 2.3)$\sigma$ which can be a significant difference. These results indicate that, depending on model, the $L_X-L_{UV}$ relation can be significantly different for QSOs at $z < 1.479$ and at $z > 1.479$.\footnote{We thank Guido Risaliti for pointing out that this could be a reflection of the inadequacy of the cosmological models we use here, and not an indication of the redshift dependence of the $L_X-L_{UV}$ relation. Also, we note that this QSO sample has been compiled from 7 different samples \citep{Lusso2020} and so is highly heterogeneous. It will be valuable to determine whether or not the heterogeneous nature of these data is related to the issues we have found in this paper. We hope that further study of the quasar measurements will resolve this issue.}$^,$ \footnote{For a possibly related issue, see \cite{Banerjee2020}.} We note that the differences are more significant in the non-flat cases. The $\Delta \gamma$ and $\Delta \beta$ values for the QSO$-z < 1.497$ and QSO comparison, and for the QSO$-z > 1.479$ and QSO comparison, listed in the bottom two-thirds of Table 5, are for illustrative purposes only, as there are correlations between the two data sets in each pair.

These results indicate that the $L_X-L_{UV}$ relation parameters can be cosmological-model dependent as well as redshift dependent. Fortunately, the $L_X-L_{UV}$ relation parameters are not model-dependent for the QSO$-z < 1.479$ data subset (Table 6), so this part of the current version of these QSO data might be a valid and useful cosmological probe. We note that, from the relevant lines in Table 3, all three data sets, QSO, QSO$-z < 1.479$, and QSO$-z > 1,479$, favor $\Omega_{m0}$ values that are not consistent with most other determinations that favor $\Omega_{m0} \sim 0.3$.\footnote{Except in the flat XCDM parametrization where QSO$-z > 1.479$ does not constrain $\Omega_{m0}$, and in the non-flat XCDM parameterization where QSO$-z < 1.479 (> 1.479)$ require $\Omega_{m0} > 0.285 (> 0.333)$ at 2$\sigma$.} However, because of the smaller number of QSOs in the $z < 1.479$ and $z > 1.479$ subsets compared to the full QSO data set, the $\Omega_{m0}$ error bars (or limits) are less restrictive for the smaller data subsets, and consequently these two $\Omega_{m0}$ determination are in less significant conflict with $\Omega_{m0} \sim 0.3$. In Fig.\ 8 we have compared the QSO$-z < 1.479$ flat $\Lambda$CDM best-fit model that has $\Omega_{m0} = 0.670$ with the $\Omega_{m0} = 0.3$ flat $\Lambda$CDM model and the Hubble diagram of the full QSO data set. This figure shows that QSOs at $z \lesssim 1.5$ are in comparatively less conflict with the $\Omega_{m0} = 0.3$ flat $\Lambda$CDM model than is the full QSO data set.

For the QSO$-z < 1.479$ data subset, from Table 6, the difference between $\delta$ values ($\Delta \delta$) from model to model is zero. The difference between $\gamma$ values ($\Delta \gamma$) ranges between (0 -- 0.16)$\sigma$ which is not statistically significant. The difference between $\beta$ values ($\Delta \beta$) ranges between (0 -- 0.19)$\sigma$ which is not statistically significant. From these results we can conclude that QSOs in the QSO$-z < 1.479$ subset are all potentially standardizable quasars. For the QSO$-z > 1.479$ data subsets, from the Table 7, the difference between $\delta$ values ($\Delta \delta$) from model to model ranges between (0--0.71)$\sigma$ which is not statistically significant. The difference between $\gamma$ values ($\Delta \gamma$) ranges between (0 -- 2.5)$\sigma$ which can be statistically significant. The difference between $\beta$ values ($\Delta \beta$) ranges between (0.02 -- 2.4)$\sigma$ which can be statistically significant. From Table 7, the QSO$-z > 1.479$ data subset appears to include QSOs that are not standard candles.

It is of interest to determine the highest redshift to which we can retain QSOs in the current compilation and still have $L_X-L_{UV}$ relation parameters that are model independent. To examine this issue, we consider three additional QSO data subsets, QSO$-z < 1.75$, QSO$-z < 2$, and QSO$-z < 2.25$ with 1313, 1534, and 1680 QSOs. Results from these analysis are listed in Tables 2, 3, and 8--10. From Table 3, for some of these data subgroups, values of $\delta$, $\gamma$, and $\beta$ are model dependent.

For the QSO$-z < 1.75$ data, from Table 8, the difference between $\delta$ values ($\Delta \delta$) from model to model ranges between (0--0.14)$\sigma$ which is not statistically significant. The difference between $\gamma$ values ($\Delta \gamma$) ranges between (0 -- 0.61)$\sigma$ which is mostly not statistically significant. The difference between $\beta$ values ($\Delta \beta$) ranges between (0.02 -- 0.68)$\sigma$ which is mostly not statistically significant. These results could be interpreted as suggesting that the QSO$-z < 1.75$ data contain QSOs that are all potentially standardizable, but this is likely to be incorrect since some of the QSO$-z < 1.75$ $\beta$ and $\gamma$ differences are significantly larger than those for QSO$-z < 1.479$ (Table 6) and these large changes have been caused by the addition of just 294 (relative to 1019) more QSOs. 

For the QSO$-z < 2$ data, from Table 9, the difference between $\delta$ values ($\Delta \delta$) from model to model ranges between (0--0.4)$\sigma$ which is not statistically significant. The difference between $\gamma$ values ($\Delta \gamma$) ranges between (0 -- 1.4)$\sigma$ which can be statistically significant. The difference between $\beta$ values ($\Delta \beta$) ranges from (0.02 -- 1.5)$\sigma$ which can be statistically significant. For QSO$-z < 2.25$ data, from Table 10, the difference between $\delta$ values ($\Delta \delta$) from model to model ranges between (0--0.53)$\sigma$ which is not statistically significant. The difference between $\gamma$ values ($\Delta \gamma$) ranges between (0 -- 2.0)$\sigma$ which can be statistically significant. The difference between $\beta$ values ($\Delta \beta$) ranges between (0.02 -- 2.1)$\sigma$ which can be statistically significant. The QSO$-z < 2$ and QSO$-z <2.25$ data subsets probably should not be used to constrain cosmological parameters.

In Figs.\ 2--7, we plot lines corresponding to parameter values for which the current cosmological expansion is unaccelerated for each model. From these figures, in spatially-flat models, the full QSO data completely favor currently decelerating cosmological expansion, strongly contradicting constraints from most other cosmological data. In non-flat models, for the full QSO data, part of the two-dimensional confidence contours lie in the currently accelerating region and part lie in the currently decelerating region, depending on cosmological model, so results from the full QSO data in non-flat models are in less conflict with the currently accelerated cosmological expansion that is favored by most other data. Similarly, the higher redshift QSO data subsets also mostly favor currently decelerating expansion in spatially-flat models, while results from spatially non-flat models are in less conflict with currently accelerated cosmological expansion. On the other hand, the lower redshift QSO data subsets, QSO$-z < 1.479$ and QSO$-z < 1.75$, results are relatively more consistent with currently accelerated expansion. It is somewhat concerning that the higher redshift QSO data, in spatially-flat models, favor currently decelerating cosmological expansion.

On the whole it appears that most QSOs in the QSO$-z < 1.479$ data subset of the \cite{Lusso2020} compilation obey an $L_X-L_{UV}$ relation that is independent of cosmological model; and that many (but not all) of the QSOs in the QSO$-z < 1.75$ data subset also obey an $L_X-L_{UV}$ relation that is independent of cosmological model. Consequently it is reasonable to use QSO$-z < 1.479$ data (as well as possibly QSO$-z < 1.75$ data) to constrain cosmological parameters. We emphasize, however, that it seems incorrect to use the higher redshift ($z \gtrsim 1.5-1.7$) part of these QSO data to constrain cosmological parameters, and it might also be better to carefully examine whether current QSO data at $z \sim 1.4-1.5$ might need to be rejected for this purpose.

\subsection{Cosmological constraints from the QSO$-z < 1.479$ and QSO$-z < 1.75$ data subsets}
\label{QSO-br}
From the analysis of the full QSO data, it appears that it includes QSOs that are not standardizable. However, QSOs in the QSO$-z < 1.479$ and QSO$-z < 1.75$ data subsets appear to be mostly standardizable and so we use these data subsets to constrain cosmological and $L_X-L_{UV}$ relation parameters. Results are given in Table 2 and 3 and shown in Fig.\ 2--7.

For the QSO$-z < 1.479$ data, the value of $\Omega_{m0}$ ranges from $0.600^{+0.340}_{-0.170}$ to $0.670^{+0.300}_{-0.130}$. The minimum value is obtained in the flat $\phi$CDM model and the maximum value in the flat $\Lambda$CDM model. For the QSO$-z < 1.75$ data, the minimum value of $\Omega_{m0}$, $> 0.306$, is for the non-flat XCDM parametrization and the maximum value of $\Omega_{m0}$, > 0.466, is for the flat $\Lambda$CDM model. These $\Omega_{m0}$ values are larger than those favored by other data, but have large uncertainties.

In the flat $\Lambda$CDM model, $\Omega_{\Lambda}$ are $0.330^{+0.130}_{-0.300}$ and $< 0.534$ using the QSO$-z < 1.479$ and QSO$-z < 1.75$ data subsets respectively. In the non-flat $\Lambda$CDM model, $\Omega_{\Lambda}$ are $0.810^{+0.810}_{-0.340}$ and $1.300^{+0.350}_{-0.099}$ for the QSO$-z < 1.479$ and QSO$-z < 1.75$ data subsets respectively.

Among all three non-flat models, for these two data subsets, the minimum value of $\Omega_{k0}$, $-1.040^{+0.210}_{-0.350}$, is obtained in the non-flat $\Lambda$CDM model using the QSO$-z < 1.75$ data and the maximum value of $\Omega_{k0}$, $-0.110^{+0.330}_{-0.330}$, is obtained in the non-flat $\phi$CDM model for the QSO$-z < 1.479$ data.\footnote{In the non-flat $\Lambda$CDM model and the non-flat XCDM parametrization, the current value of the dark energy density parameter is determined from the chosen values of $\Omega_{m0}$ and $\Omega_{k0}$. So, there is no restriction on the dark energy density parameter. In the non-flat $\phi$CDM model, the value of the dark energy density parameter $\Omega_{\phi}(z, \alpha)$ is determined numerically by solving the dynamical equations and it's value always lies in the range $0 \leq \Omega_{\phi}(0, \alpha) \leq 1$. In the non-flat $\phi$CDM model plots, this restriction on $\Omega_{\phi}(0,\alpha)$ can be seen in the $\Omega_{m0}-\Omega_{k0}$ sub-panel in the form of straight line boundaries.} From the Table 3 results we see that these QSO data subsets tend to favor closed geometries.

In the flat XCDM parameterization, $\omega_X$ are determined to be $< -0.137$ and $< -0.013$ using QSO$-z < 1.479$ and QSO$-z < 1.75$ data. In the non-flat XCDM parameterization,
$\omega_X$ are  $-0.230^{+0.520}_{-0.450}$ and $-0.740^{+0.560}_{-0.440}$ for the QSO$-z < 1.479$ and QSO$-z < 1.75$ data subsets. In the flat $\phi$CDM model, the values of the scalar potential parameter $\alpha$ are $5.200^{+3.800}_{-2.500}$ and $5.300^{+4.200}_{-1.900}$ using the QSO$-z<1.479$ and QSO$-z<1.75$ data subsets respectively. In the non-flat $\phi$CDM model, QSO$-z<1.479$ and QSO$-z<1.75$ data cannot constrain $\alpha$.

From Table 2, for the QSO$-z<1.479$ data subset, from both the $AIC$ and $BIC$ values, the most favored model is the flat $\Lambda$CDM model and the least favored model is the non-flat $\phi$CDM model. For QSO$-z<1.75$ data, from the $AIC$ values, the most favored model is the non-flat $\Lambda$CDM model and the least favored model is the flat $\phi$CDM model, and from the $BIC$ values, the most favored model is the flat $\Lambda$CDM model and the least favored model is the non-flat $\phi$CDM model.

\subsection{Cosmological constraints from the BAO + $H(z)$ data}
\label{BAO+H(z)}
The constraints obtained using the BAO + $H(z)$ data are given in Table 2 and 3 and the one-dimensional likelihoods and two-dimensional constraint contours are plotted in Figs.\ 2--7.

From Table 3, the minimum value of $\Omega_b h^2$, $0.024^{+0.003}_{-0.003}$, is obtained in the flat $\Lambda$CDM model while the maximum value of $\Omega_b h^2$, $0.032^{+0.006}_{-0.003}$, is obtained in the flat $\phi$CDM model. The minimum value of $\Omega_c h^2$, $0.081^{+0.017}_{-0.017}$, is
for the flat $\phi$CDM model and the maximum value of $\Omega_c h^2$, $0.119^{+0.008}_{-0.008}$, is for the flat $\Lambda$CDM model. The minimum value of $\Omega_{m0}$, $0.266^{+0.023}_{-0.023}$, is obtained in the flat $\phi$CDM model while the maximum value of $\Omega_{m0}$, $0.299^{+0.015}_{-0.017}$, is obtained in the flat $\Lambda$CDM model. These $\Omega_{m0}$ values are reasonably consistent with those determined using other data.

Form Table 3, for the BAO + $H(z)$ data, the value of $H_0$ is determined to lie in the range $65.100^{+2.100}_{-2.100}$ to $69.300^{+1.800}_{-1.800}$ ${\rm km}\hspace{1mm}{\rm s}^{-1}{\rm Mpc}^{-1}$. The minimum value is for the spatially-flat $\phi$CDM model while the maximum value is in the spatially-flat $\Lambda$CDM model. These values are more consistent with the  \cite{Plank2018} and median statistics \citep{chen3} results than with the larger local expansion rate value of \cite{Riess2016}.\footnote{Other local expansion rate determinations have slightly lower central values with slightly larger error bars \citep{Rigault2015,Zhang2017,Dhaw,Fernandez2018,Freedman2020,Rameez2019,Breuval2020,Efstathiou2020,Khetan2020}. Our $H_0$ measurements are consistent with earlier median statistics estimates \citep{Gott2001, chenetal2003} and a number of recent $H_0$ measurements made using a variety of techniques \citep{chen5, DESb, Gomez2018, Plank2018, Zhang2018a, D, Martinelli2019, Cuceu2019, Zeng2019, schon2019,  LinI2019,Blum2020,Lyu2020, phil2020,Zhang2020,Birrer2020,Denzel2020,Pogosian2020,Boruah2020,Kim2020}.}

The value of $\Omega_{\Lambda}$ in the flat and non-flat $\Lambda$CDM model is measured to be $0.701^{+0.017}_{0.015}$ and $0.667^{+0.093}_{-0.081}$ respectively.

In the non-flat $\Lambda$CDM model, the value of $\Omega_{k0}$ is found to be $-0.014 \pm 0.075$ while in the non-flat XCDM parametrization and $\phi$CDM model, the values of $\Omega_{k0}$ are $-0.120 \pm 0.130$ and $-0.080 \pm 0.100$ respectively. These are consistent with flat spatial hypersurfaces but do not rule out mildly curved geometry.

In the flat (non-flat) XCDM parametrization, $\omega_X$ is measured to be $-0.744^{+0.140}_{-0.097}(-0.693^{+0.130}_{-0.077})$. In the flat (non-flat) $\phi$CDM model, the scalar potential enrgy density parameter $(\alpha)$ is found to be $1.530^{+0.620}_{-0.850}(1.660^{+0.670}_{-0.830})$. These values of $\omega_X$ and $\alpha$ favor dynamical dark enery models over the cosmological constant at a statistical significance of between (1.8-4)$\sigma$.

From Table 2, from both the $AIC$ and $BIC$ values, the most favored model is the spatially-flat $\phi$CDM model while non-flat $\Lambda$CDM model is the least favored.

\subsection{Cosmological constraints from the QSO + BAO + $H(z)$ data}
\label{QSO+BAO+H(z)}
The QSO$-z < 1.479$ (and slightly less so, the QSO$-z < 1.75$) data constraints are consistent with the BAO + $H(z)$ constraints. So, it is reasonable to perform joint analyses of these data subsets in combination with the BAO + $H(z)$ data. The constraints obtained using QSO$-z < 1.497$ + BAO + $H(z)$ and QSO$-z < 1.75$ + BAO + $H(z)$ data are given in the corresponding lines of Tables 2 and 3 while the one-dimensional likelihoods and two-dimensional contours are shown in Figs.\ 9--11. These QSO data subsets do not significantly tighten the BAO + $H(z)$ data contours and so do not significantly alter the cosmological parameter values determined from the BAO + $H(z)$ data. 

From Table 2, for the QSO$-z < 1.497$ + BAO + $H(z)$ data set,  from the $AIC$ values, the most-favored model is the spatially-flat $\phi$CDM model while the least favored is the non-flat $\Lambda$CDM model, and from the $BIC$ values, the most-favored model is the spatially-flat $\Lambda$CDM model and the least favored is the non-flat $\phi$CDM model. For the QSO$-z < 1.75$ + BAO + $H(z)$ data set, from the $AIC$ values, the most-favored model is the non-flat $\phi$CDM model while the least favored is the non-flat $\Lambda$CDM model, and from the $BIC$ values, the most-favored model is the spatially-flat $\Lambda$CDM model and the least favored is the non-flat $\Lambda$CDM model.

\section{Conclusion}
\label{con}
The first large compilation and detailed study of ($\sim 800$) quasars as a cosmological probe was described in \cite{Risaliti2015}. Cosmological constraints obtained using these data were in agreement with those obtained using other cosmological probes \citep{Risaliti2015,Khadka2020a}. In 2019, \cite{Risaliti2019} updated their data set to include $\sim 1600$ QSO measurements. The Hubble diagram of this updated data set was somewhat inconsistent with that of a flat $\Lambda$CDM model with $\Omega_{m0} = 0.3$; basically these updated QSO data favored a value of $\Omega_{m0}$ larger than 0.3, but this tension was mild \citep{Khadka2020b}.

More recently, in 2020, \cite{Lusso2020} released an updated, larger, QSO data set of 2038 higher-quality (out of 2421) measurements. For the full 2038 QSO data set, we find that the $L_X-L_{UV}$ relation parameter values depend on the cosmological model assumed in the determination of these parameters. If instead we use the lower redshift part of this compilation, with $z \lesssim 1.5-1.7$ (and with $\sim 1000-1300$ QSO measurements), then the resulting $L_X-L_{UV}$ relation parameter values are almost independent of the assumed cosmological model and so these smaller, lower redshift, QSO data subsets can be used as cosmological probes. However they still favor higher $\Omega_{m0}$ values than do most other cosmological probes. While the cosmological constraints from these lower redshift, smaller, QSO data subsets are consistent with those that follow from the BAO + $H(z)$ data (which is not true of the cosmological constraints that follow from the full (2038) QSO data set), these smaller QSO data subsets do not significantly alter the BAO + $H(z)$ cosmological constraints when they are jointly analysed with the BAO + $H(z)$ data.

While our results here indicate the QSOs with $z \sim 1.5-1.7$ and larger in the latest \cite{Lusso2020} compilation do not obey the $L_X-L_{UV}$ relation in a cosmological-model independent manner, a more careful study is needed to discover the reason(s) for this. If higher redshift quasars, with $z \sim 2-8$, can be standardized, they should be very valuable cosmological probes.

\section{ACKNOWLEDGEMENTS}
We thank Elisabeta Lusso, Guido Risaliti, and Salvatore Capozziello for providing us with the QSO data and for valuable comments. We thank Javier De Cruz, Adria G{\'o}mez-Valent, and Chan-Gyung Park for useful suggestions about the MontePython and CLASS codes. We are grateful to the Beocat Research Cluster at Kansas State University team. This research was supported in part by DOE grant DE-SC0011840.

\section*{Data availability}
The data underlying this article were provided to us by the authors of \cite{Lusso2020}. These data will be shared on request to the corresponding author with the permission of the authors of \cite{Lusso2020}.

\begin{figure*}
\begin{multicols}{2}
    \includegraphics[width=\linewidth]{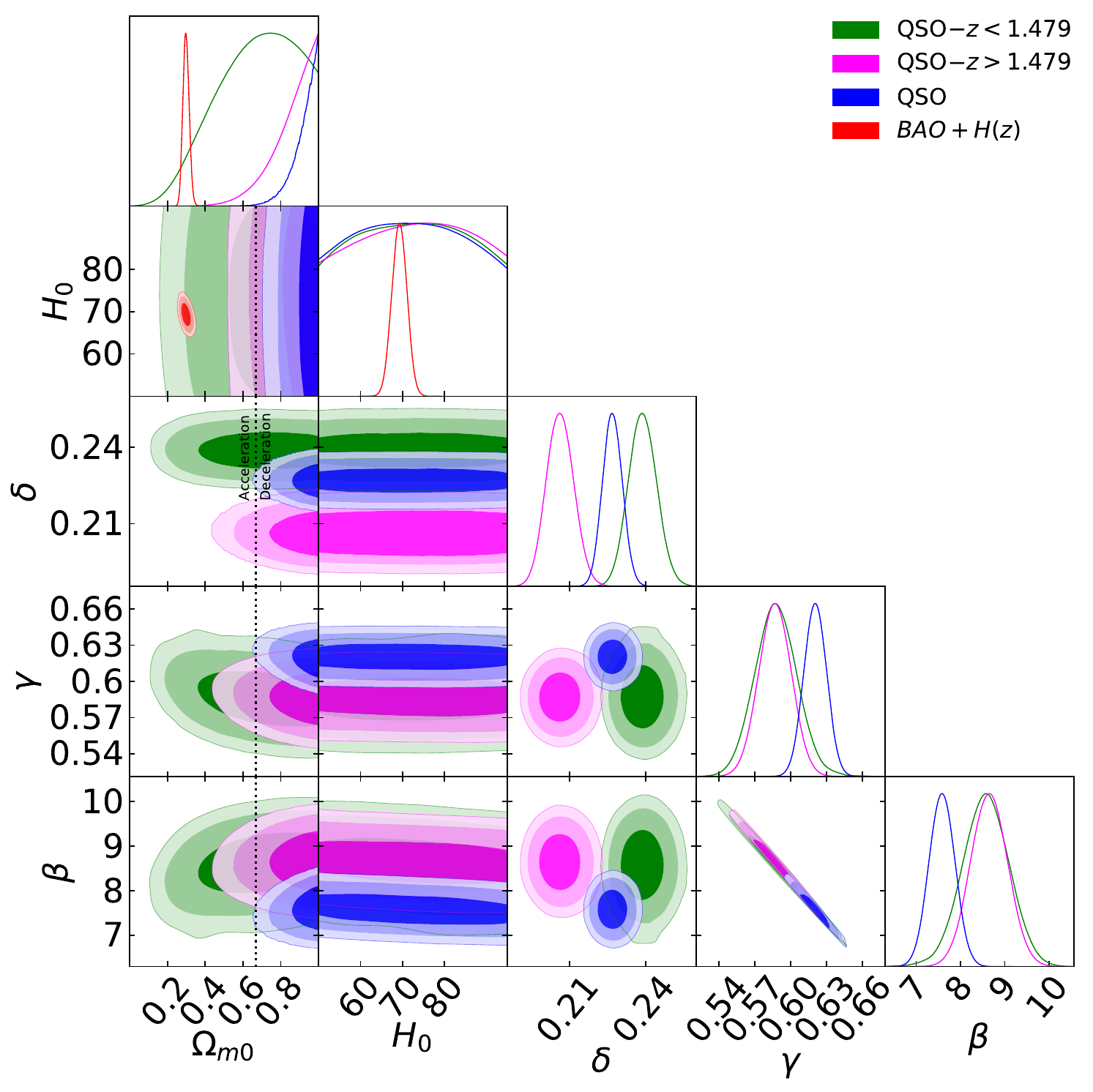}\par
    \includegraphics[width=\linewidth]{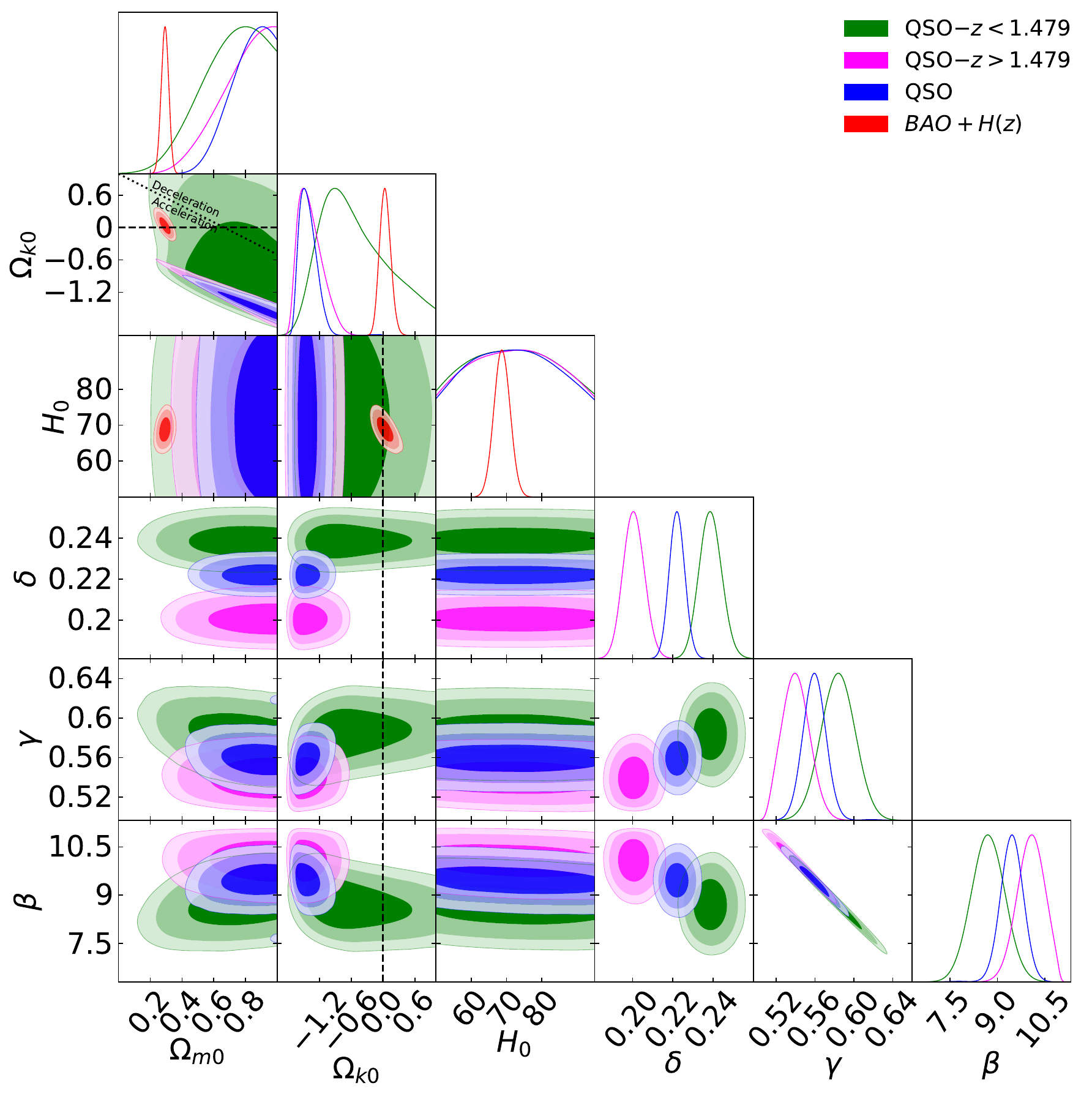}\par
\end{multicols}
\caption{One-dimensional likelihood distributions and two-dimensional contours at 1$\sigma$, 2$\sigma$, and 3$\sigma$ confidence levels using QSO$-z<1.497$ (green), QSO$-z>1.497$ (magenta), QSO (blue),  and BAO + $H(z)$ (red) data for all free parameters. Left panel shows the flat $\Lambda$CDM model. The black dotted vertical lines are the zero acceleration lines with currently accelerated cosmological expansion occurring to the left of the lines. Right panel shows the non-flat $\Lambda$CDM model. The black dotted sloping line in the $\Omega_{k0}-\Omega_{m0}$ panel is the zero acceleration line with currently accelerated cosmological expansion occurring to the lower left of the line. The black dashed horizontal or vertical line in the $\Omega_{k0}$ subpanels correspond to $\Omega_{k0} = 0$.}
\label{fig:Eiso-Ep}
\end{figure*}





\begin{figure*}
\begin{multicols}{2}
    \includegraphics[width=\linewidth]{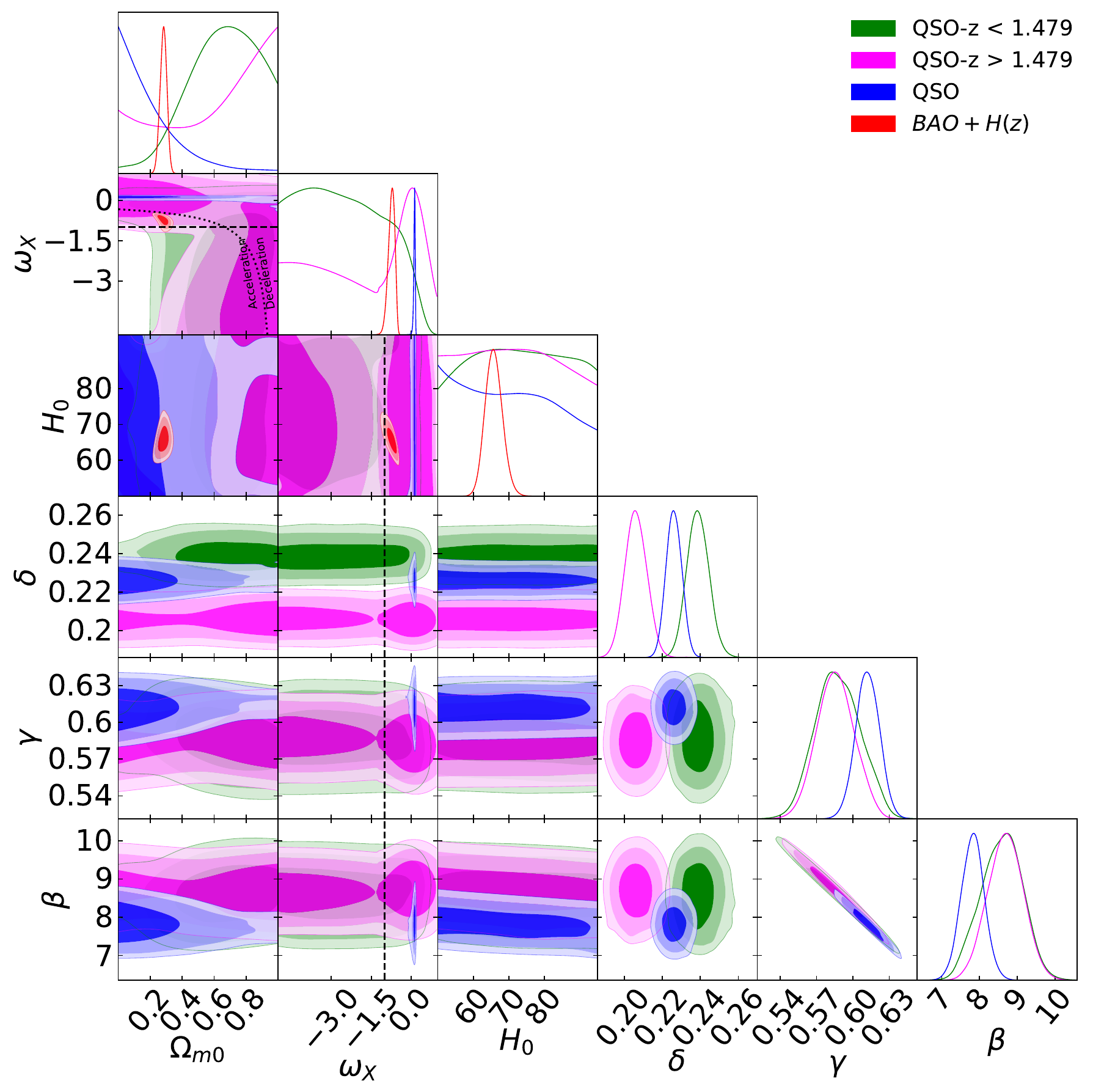}\par
    \includegraphics[width=\linewidth]{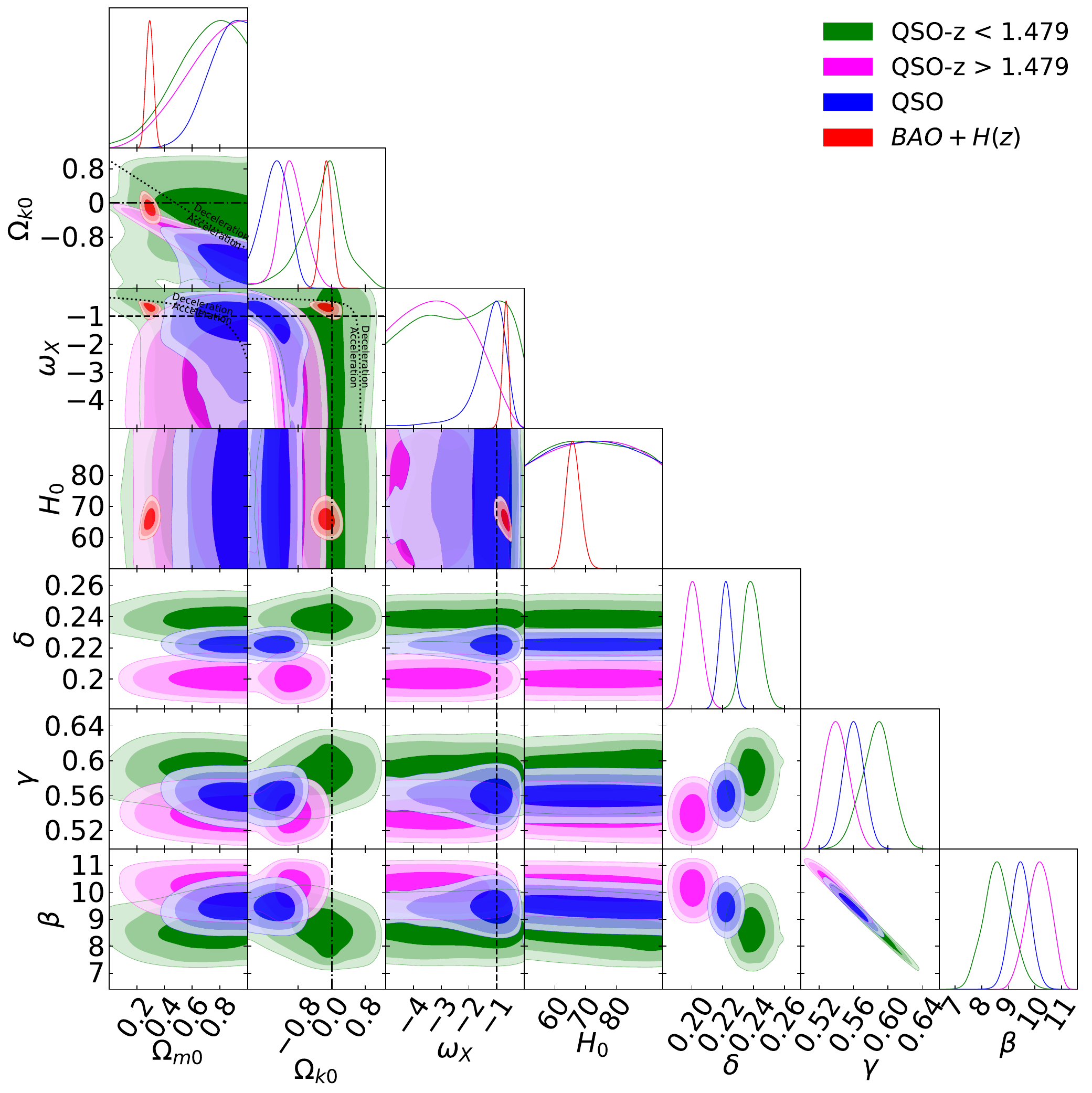}\par
\end{multicols}
\caption{One-dimensional likelihood distributions and two-dimensional contours at 1$\sigma$, 2$\sigma$, and 3$\sigma$ confidence levels using QSO$-z<1.497$ (green), QSO$-z>1.497$ (magenta), QSO (blue),  and BAO + $H(z)$ (red) data for all free parameters. Left panel shows the flat XCDM parametrization. The black dotted curved line in the $\omega_X-\Omega_{m0}$ panel is the zero acceleration line with currently accelerated cosmological expansion occurring below the line and the black dashed straight lines correspond to the $\omega_X = -1$ $\Lambda$CDM model. Right panel shows the non-flat XCDM parametrization. The black dotted lines in the $\Omega_{k0}-\Omega_{m0}$, $\omega_X-\Omega_{m0}$, and $\omega_X-\Omega_{k0}$ panels are the zero acceleration lines with currently accelerated cosmological expansion occurring below the lines. Each of the three lines is computed with the third parameter set to the BAO + $H(z)$ data best-fit value of Table 2. The black dashed straight lines correspond to the $\omega_x = -1$ $\Lambda$CDM model. The black dotted-dashed straight lines correspond to $\Omega_{k0} = 0$.}
\label{fig:Eiso-Ep}
\end{figure*}

\begin{figure*}
\begin{multicols}{2}
    \includegraphics[width=\linewidth]{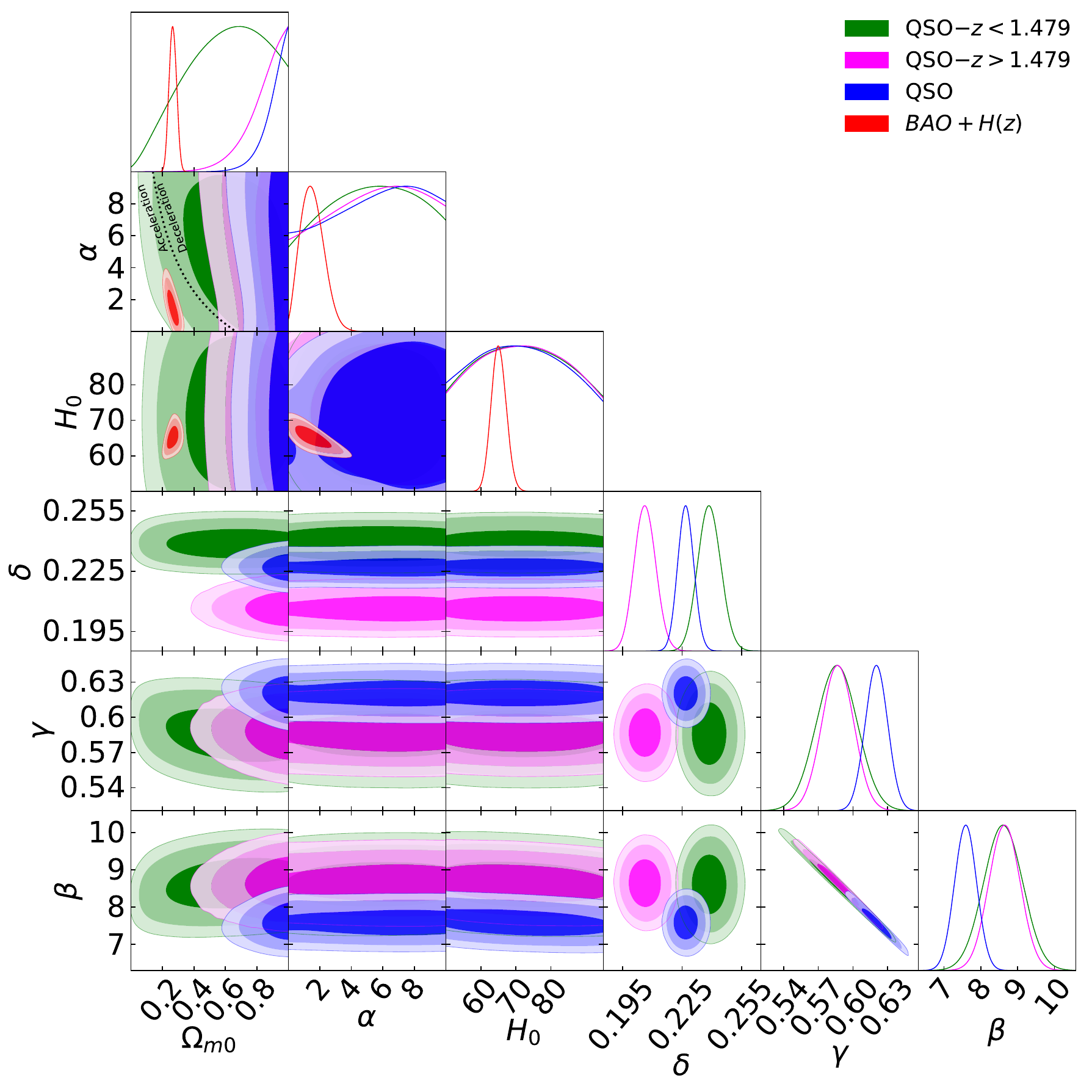}\par
    \includegraphics[width=\linewidth]{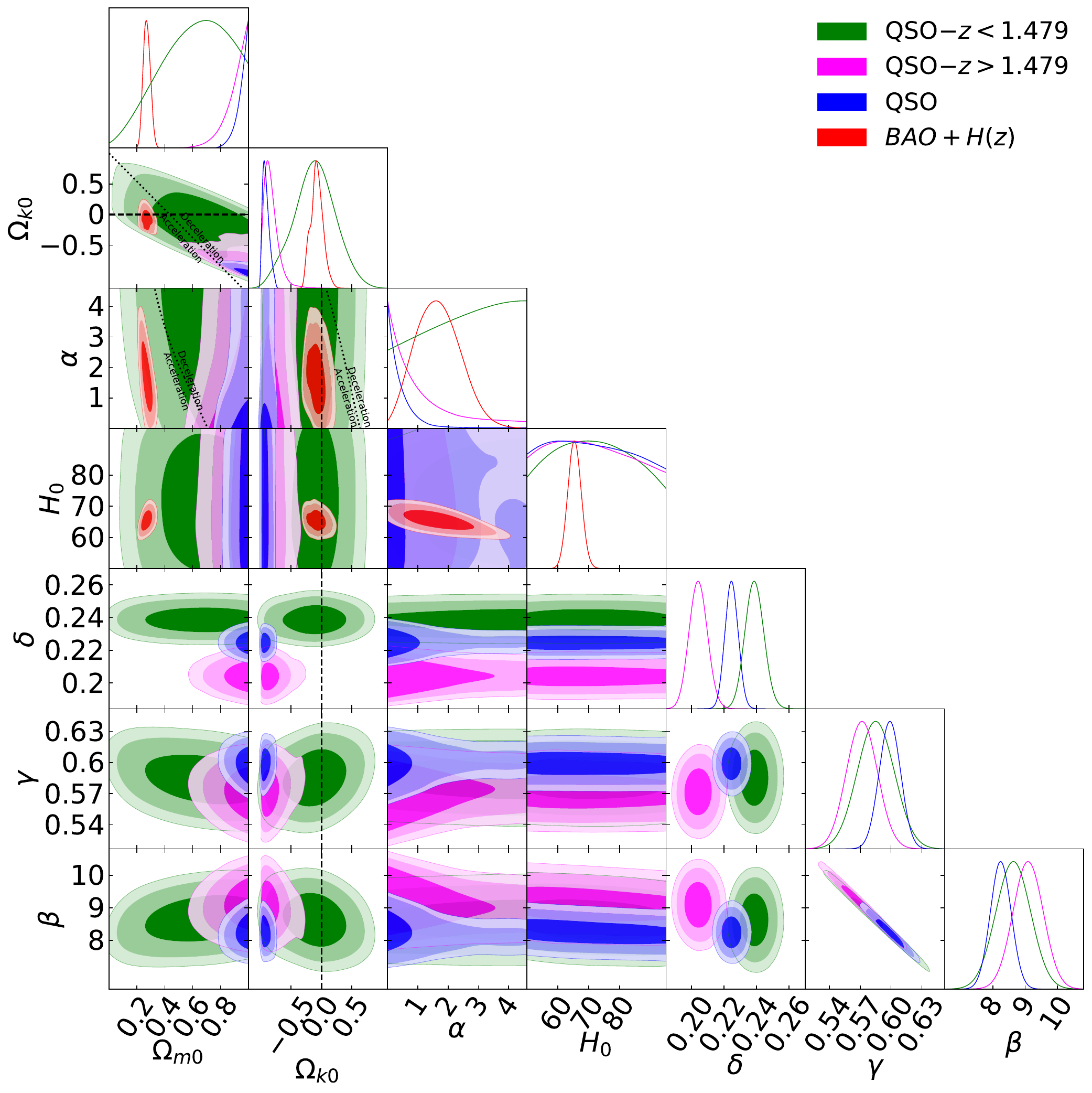}\par
\end{multicols}
\caption{One-dimensional likelihood distributions and two-dimensional contours at 1$\sigma$, 2$\sigma$, and 3$\sigma$ confidence levels using QSO$-z<1.497$ (green), QSO$-z>1.497$ (magenta), QSO (blue),  and BAO + $H(z)$ (red) data for all free parameters. The $\alpha = 0$ axes correspond to the $\Lambda$CDM model. Left panel shows the flat $\phi$CDM model. The black dotted curved line in the $\alpha - \Omega_{m0}$ panel is the zero acceleration line with currently accelerated cosmological expansion occurring to the left of the line. Right panel shows the non-flat $\phi$CDM model. The black dotted lines in the $\Omega_{k0}-\Omega_{m0}$, $\alpha-\Omega_{m0}$, and $\alpha-\Omega_{k0}$ panels are the zero acceleration lines with currently accelerated cosmological expansion occurring below the lines. Each of the three lines is computed with the third parameter set to the BAO + $H(z)$ data best-fit value of Table 2. The black dashed straight lines correspond to $\Omega_{k0} = 0$.}
\label{fig:Eiso-Ep}
\end{figure*}

\begin{figure*}
\begin{multicols}{2}
    \includegraphics[width=\linewidth]{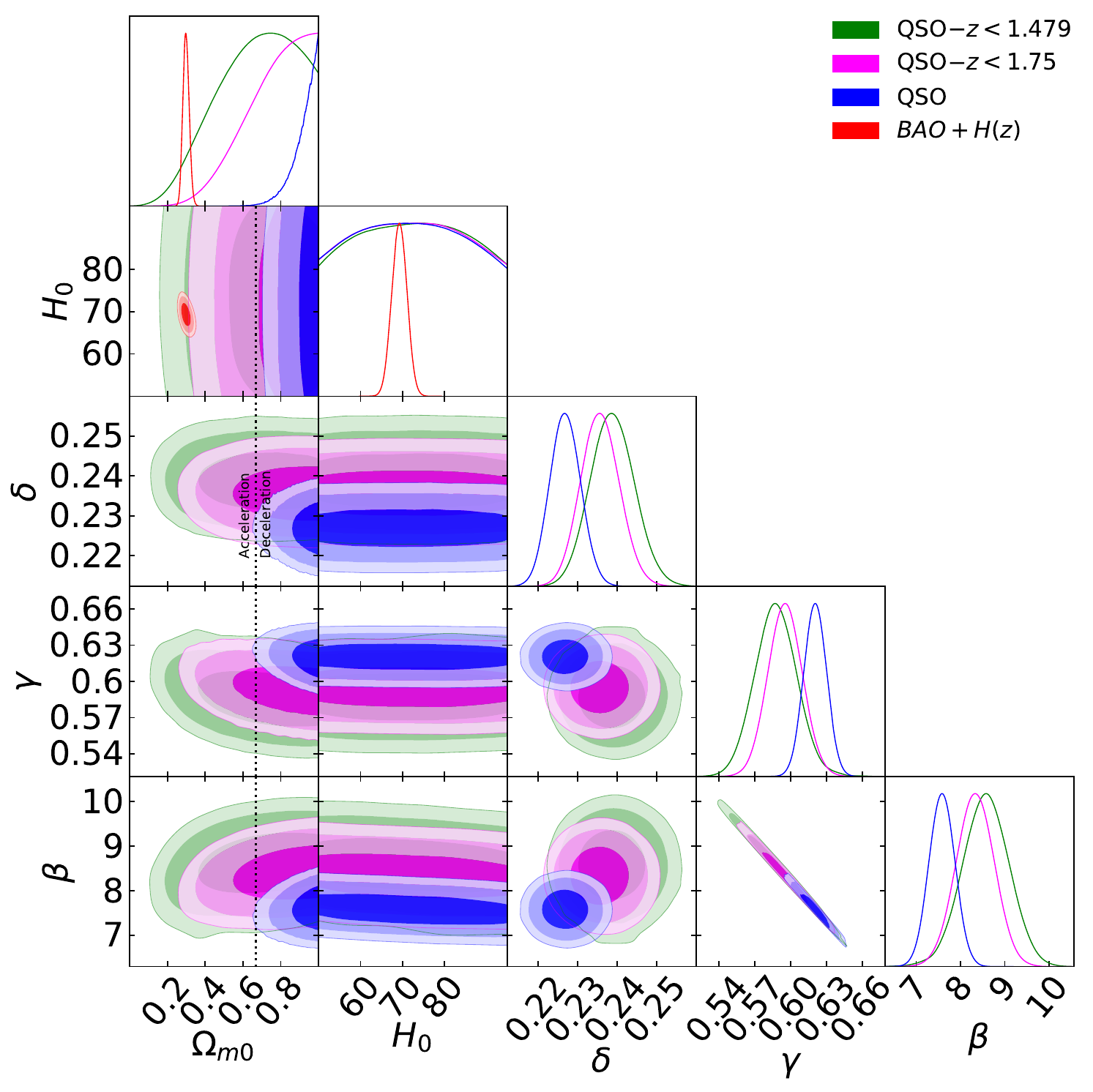}\par
    \includegraphics[width=\linewidth]{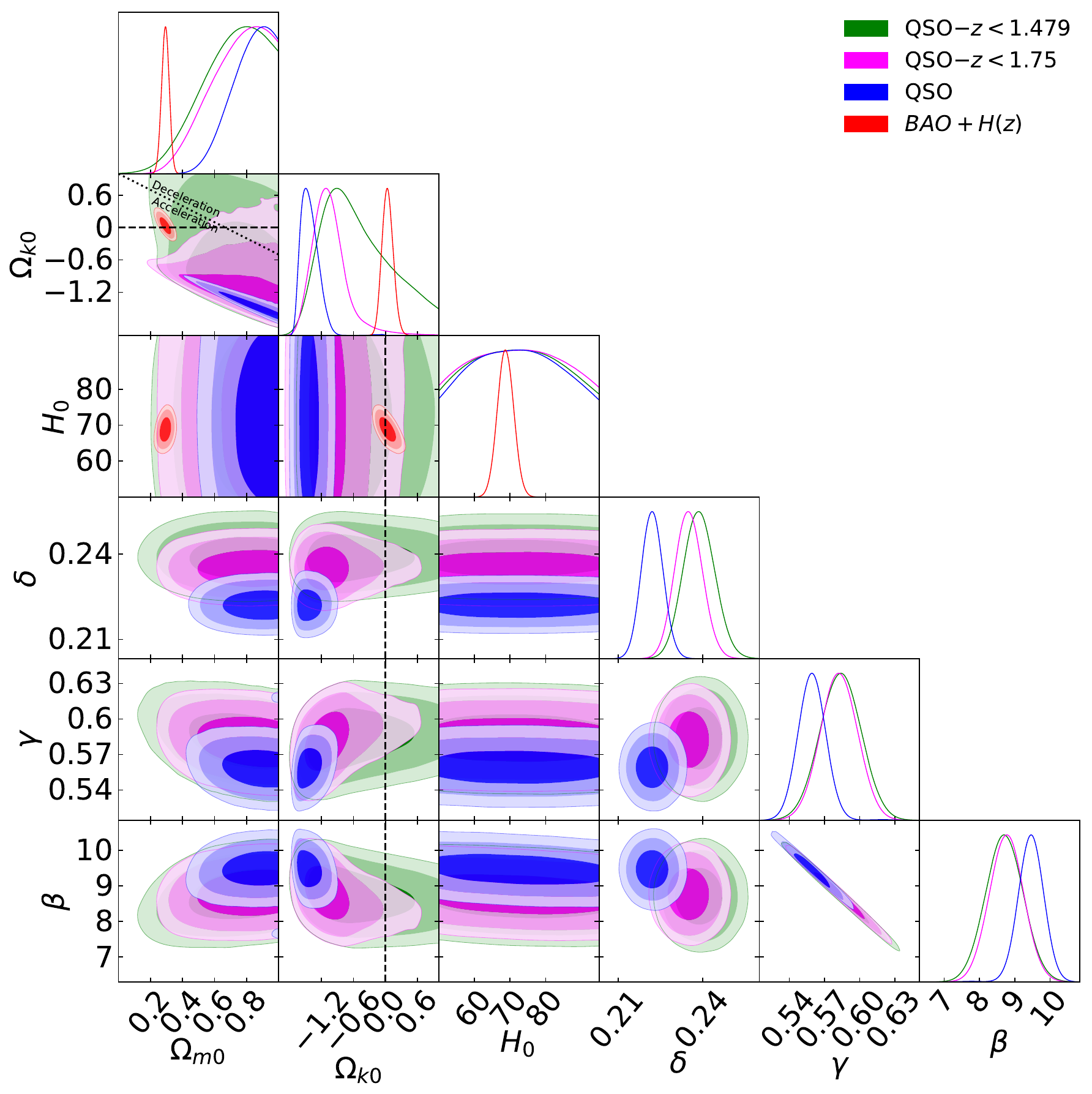}\par
\end{multicols}
\caption{One-dimensional likelihood distributions and two-dimensional contours at 1$\sigma$, 2$\sigma$, and 3$\sigma$ confidence levels using QSO$-z<1.497$ (green), QSO$-z<1.75$ (magenta), QSO (blue),  and BAO + $H(z)$ (red) data for all free parameters. Left panel shows the flat $\Lambda$CDM model. The black dotted vertical lines are the zero acceleration lines with currently accelerated cosmological expansion occurring to the left of the lines. Right panel shows the non-flat $\Lambda$CDM model. The black dotted sloping line in the $\Omega_{k0}-\Omega_{m0}$ panel is the zero acceleration line with currently accelerated cosmological expansion occurring to the lower left of the line. The black dashed horizontal or vertical line in the $\Omega_{k0}$ subpanels correspond to $\Omega_{k0} = 0$.}
\label{fig:Eiso-Ep}
\end{figure*}

\begin{figure*}
\begin{multicols}{2}
    \includegraphics[width=\linewidth]{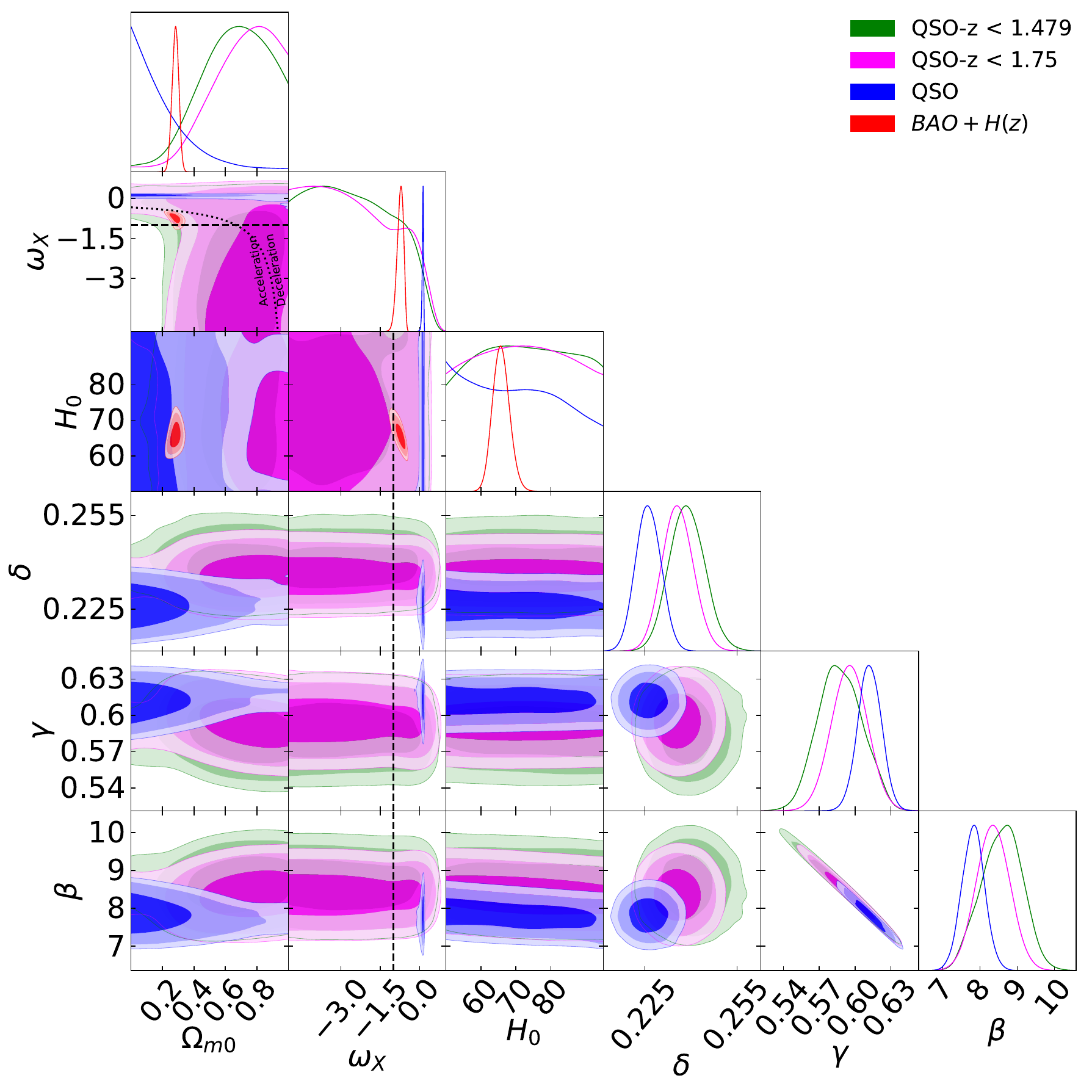}\par
    \includegraphics[width=\linewidth]{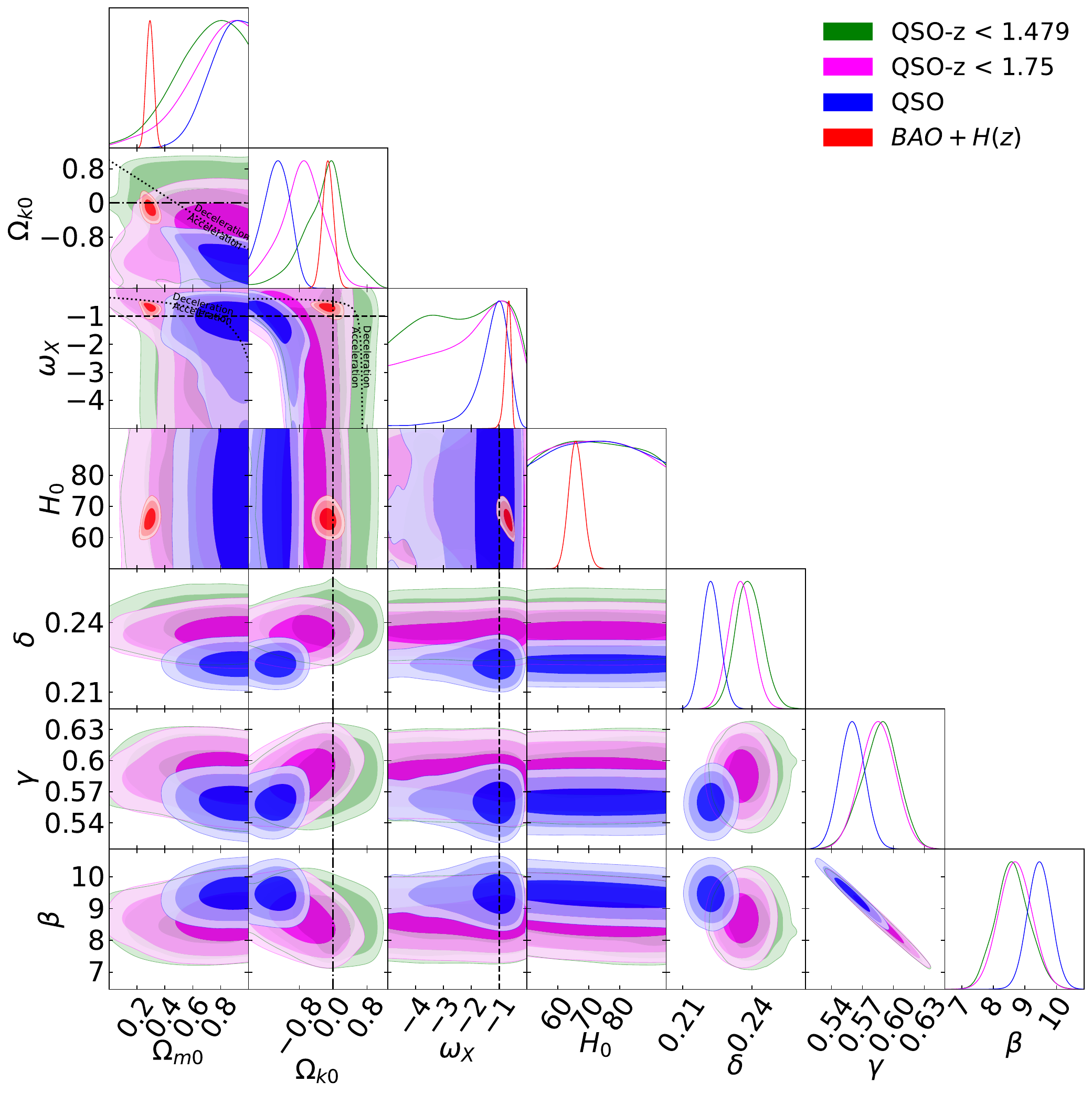}\par
\end{multicols}
\caption{One-dimensional likelihood distributions and two-dimensional contours at 1$\sigma$, 2$\sigma$, and 3$\sigma$ confidence levels using QSO$-z<1.497$ (green), QSO$-z<1.75$ (magenta), QSO (blue),  and BAO + $H(z)$ (red) data for all free parameters. Left panel shows the flat XCDM parametrization. The black dotted curved line in the $\omega_X-\Omega_{m0}$ panel is the zero acceleration line with currently accelerated cosmological expansion occurring below the line and the black dashed straight lines correspond to the $\omega_X = -1$ $\Lambda$CDM model. Right panel shows the non-flat XCDM parametrization. The black dotted lines in the $\Omega_{k0}-\Omega_{m0}$, $\omega_X-\Omega_{m0}$, and $\omega_X-\Omega_{k0}$ panels are the zero acceleration lines with currently accelerated cosmological expansion occurring below the lines. Each of the three lines is computed with the third parameter set to the BAO + $H(z)$ data best-fit value of Table 2. The black dashed straight lines correspond to the $\omega_x = -1$ $\Lambda$CDM model. The black dotted-dashed straight lines correspond to $\Omega_{k0} = 0$.}
\label{fig:Eiso-Ep}
\end{figure*}

\begin{figure*}
\begin{multicols}{2}
    \includegraphics[width=\linewidth]{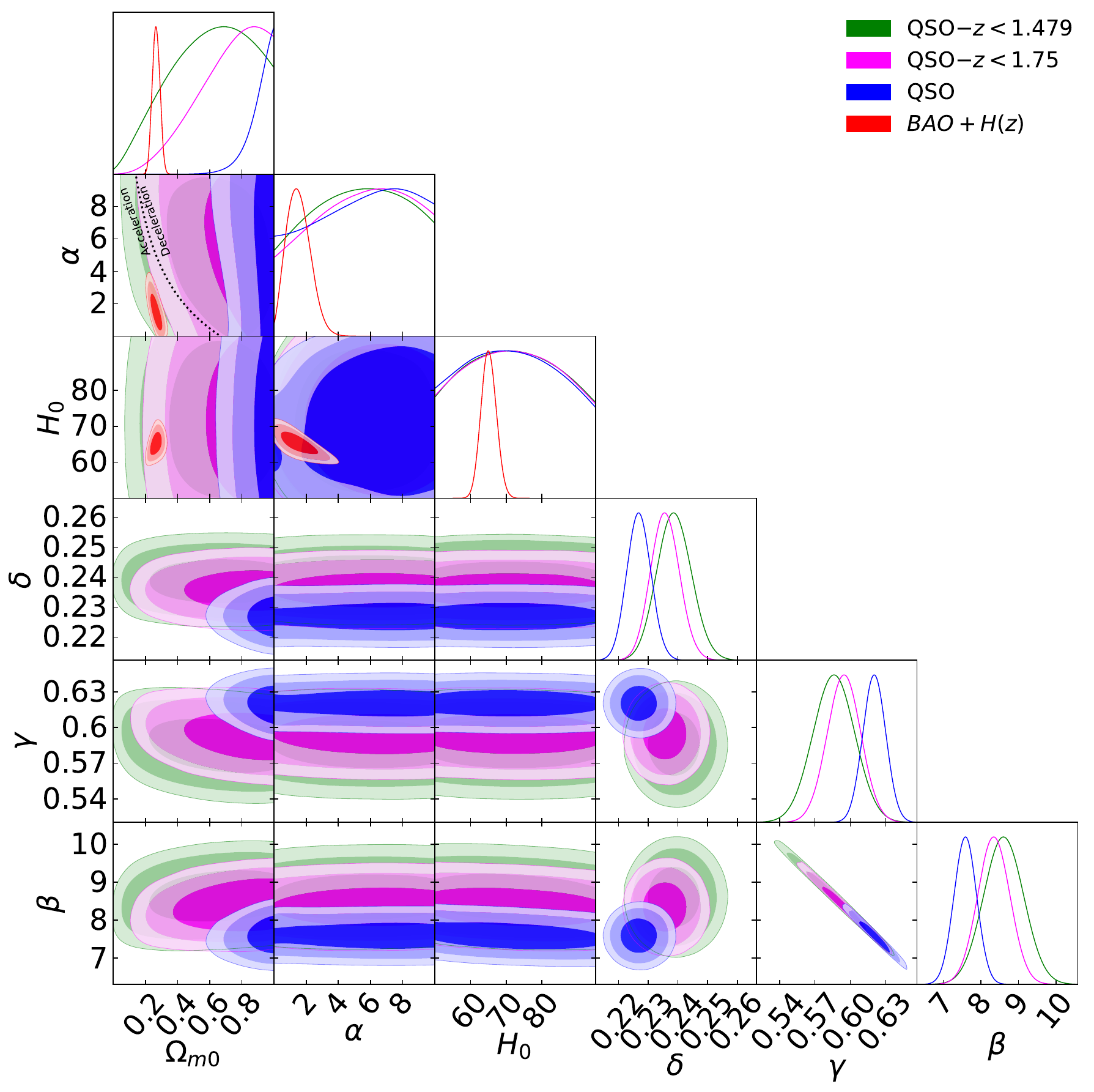}\par
    \includegraphics[width=\linewidth]{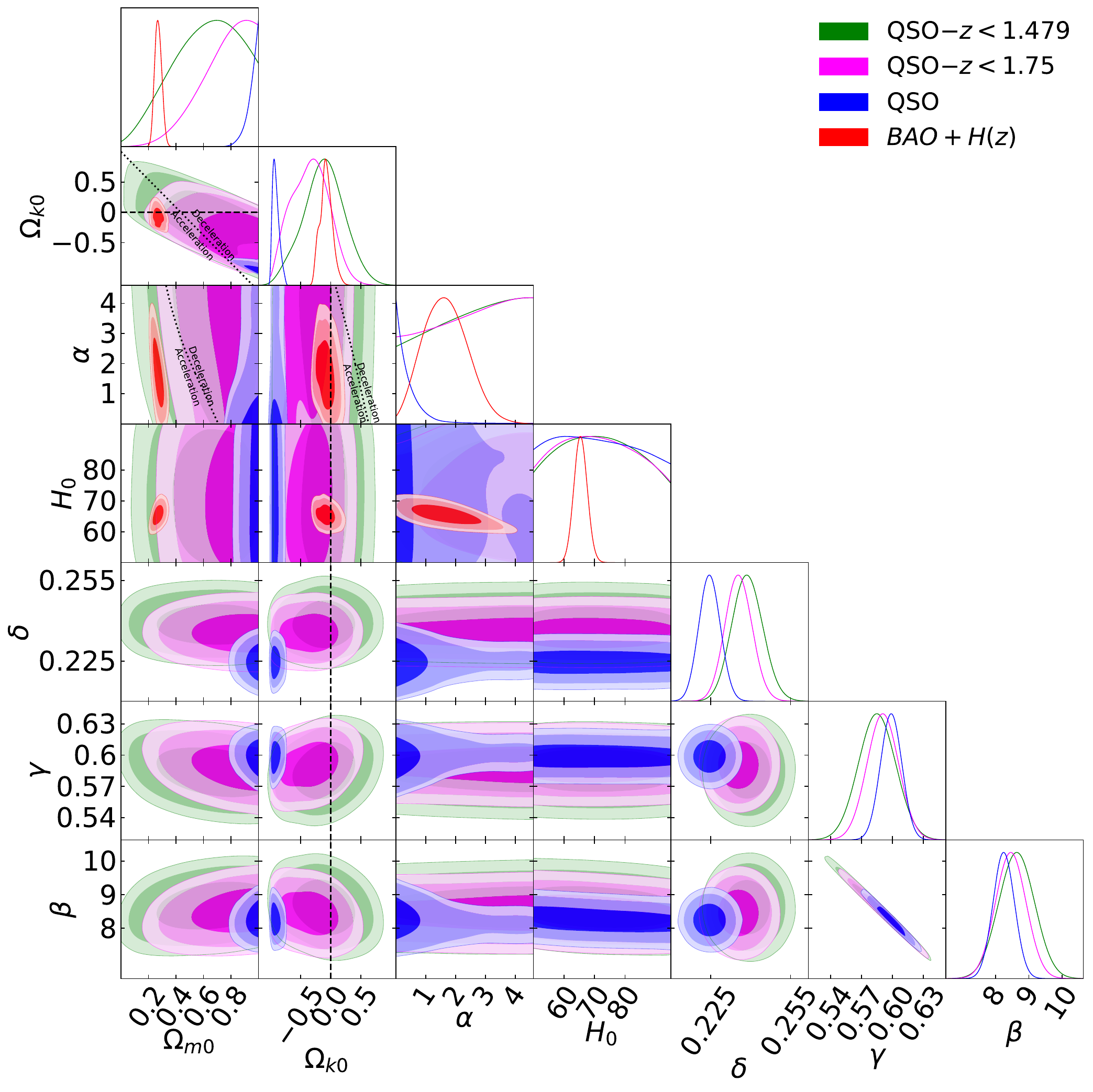}\par
\end{multicols}
\caption{One-dimensional likelihood distributions and two-dimensional contours at 1$\sigma$, 2$\sigma$, and 3$\sigma$ confidence levels using QSO$-z<1.497$ (green), QSO$-z<1.75$ (magenta), QSO (blue),  and BAO + $H(z)$ (red) data for all free parameters. The $\alpha = 0$ axes correspond to the $\Lambda$CDM model. Left panel shows the flat $\phi$CDM model. The black dotted curved line in the $\alpha - \Omega_{m0}$ panel is the zero acceleration line with currently accelerated cosmological expansion occurring to the left of the line. Right panel shows the non-flat $\phi$CDM model. The black dotted lines in the $\Omega_{k0}-\Omega_{m0}$, $\alpha-\Omega_{m0}$, and $\alpha-\Omega_{k0}$ panels are the zero acceleration lines with currently accelerated cosmological expansion occurring below the lines. Each of the three lines is computed with the third parameter set to the BAO + $H(z)$ data best-fit value of Table 2. The black dashed straight lines correspond to $\Omega_{k0} = 0$.}
\label{fig:Eiso-Ep}
\end{figure*}

\begin{figure*}
    \includegraphics[width=0.85\textwidth]{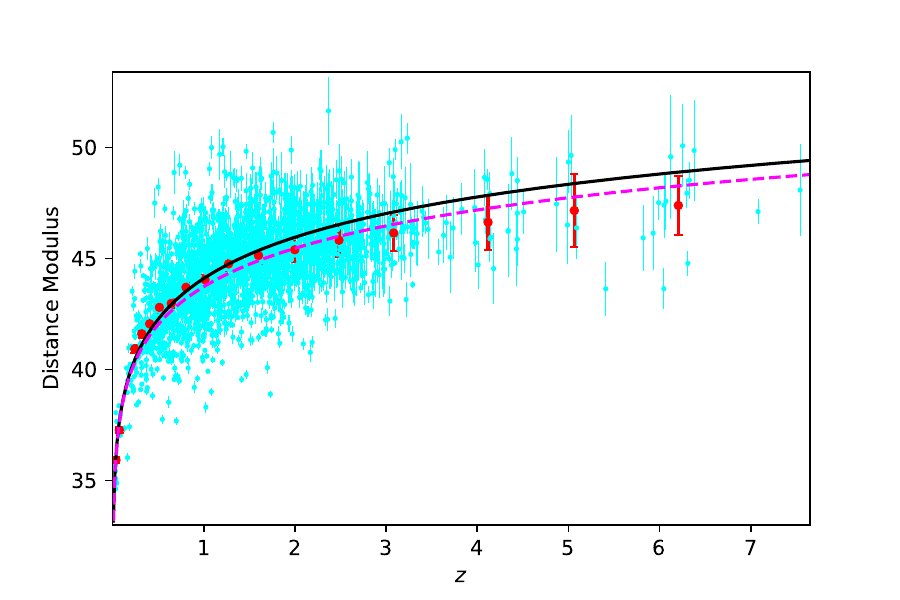}\par
\caption{Hubble diagram of quasars. Magenta dashed line is the best-fit flat $\Lambda$CDM model with $\om$ = 0.670 from the QSO-$z < 1.479$ QSO data. Cyan points are the determined QSO distance moduli and uncertainties \citep{Lusso2020} and red points are the means and uncertainties of these distance moduli, in narrow redshift bins which roughly represent a fifth order cosmographic fit of the whole QSO data \citep{Lusso2020}. The black solid line shows a flat $\Lambda$CDM model with $\om$ = 0.30.}
\label{fig: A Hubble diagram of quasar}
\end{figure*}

\begin{figure*}
\begin{multicols}{2}
    \includegraphics[width=\linewidth]{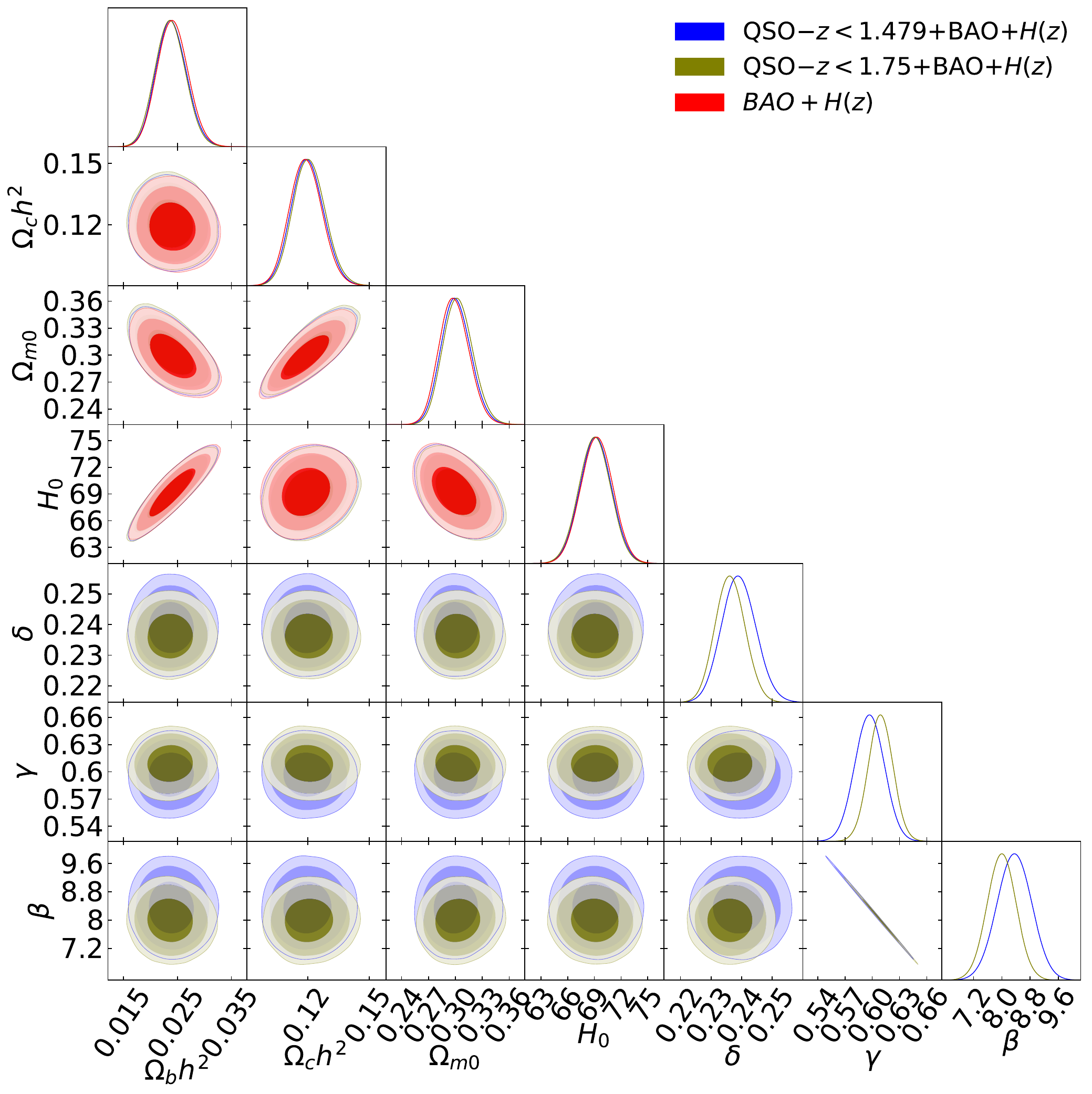}\par
    \includegraphics[width=\linewidth]{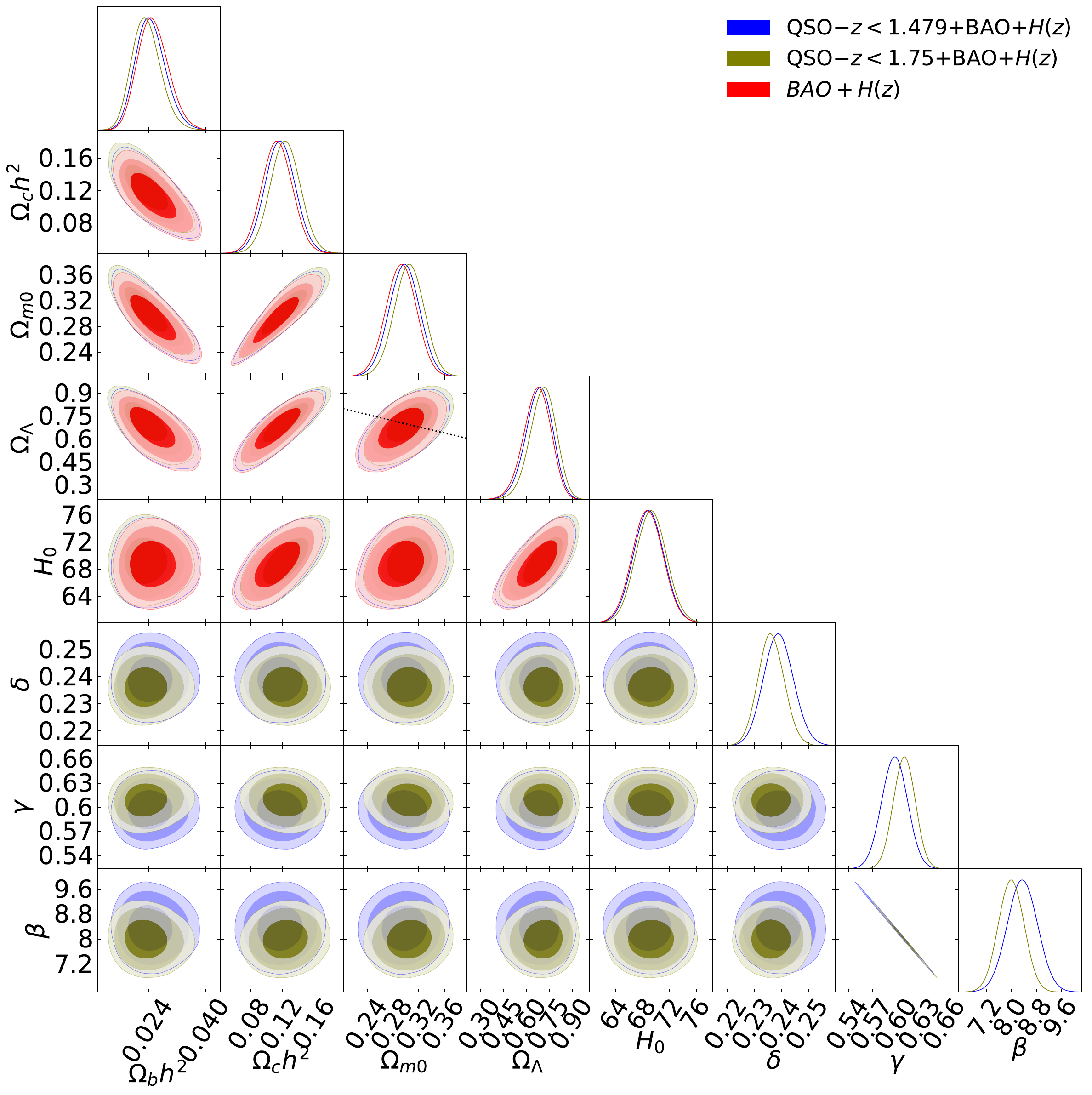}\par
\end{multicols}
\caption{One-dimensional likelihood distributions and two-dimensional contours at 1$\sigma$, 2$\sigma$, and 3$\sigma$ confidence levels using QSO$-z<1.497$ + BAO + $H(z)$ (blue), QSO$-z<1.75$ + BAO + $H(z)$ (olive), BAO + $H(z)$ (red) data for all free parameters. Left panel shows the flat $\Lambda$CDM model and right panel shows the non-flat $\Lambda$CDM model. Dotted sloping line in the $\Omega_{m0}-\Omega_{\Lambda}$ subpanel in the right panel is the $\Omega_{k0} = 0$ line.}
\label{fig:Eiso-Ep}
\end{figure*}

\begin{figure*}
\begin{multicols}{2}
    \includegraphics[width=\linewidth]{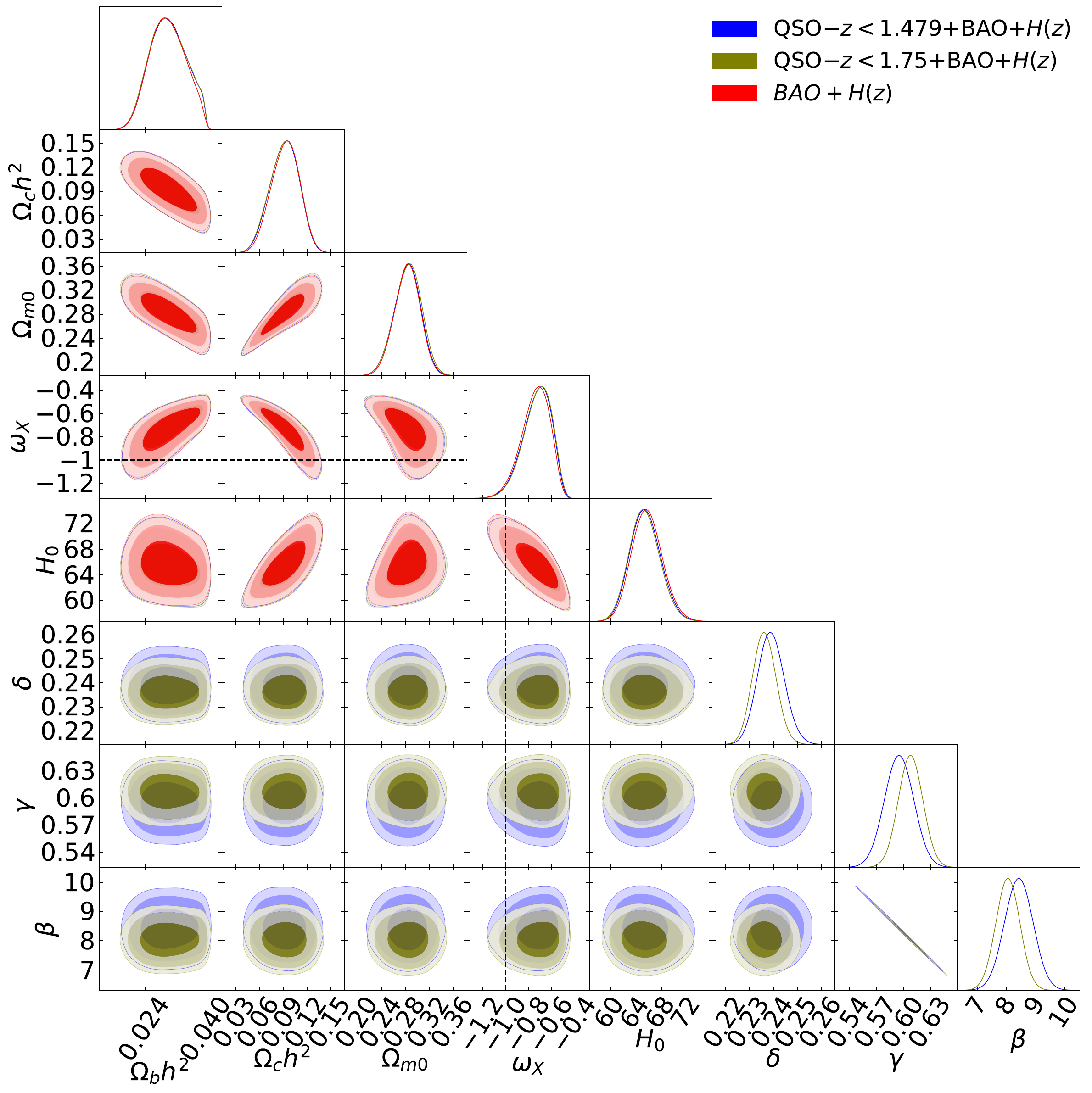}\par
    \includegraphics[width=\linewidth]{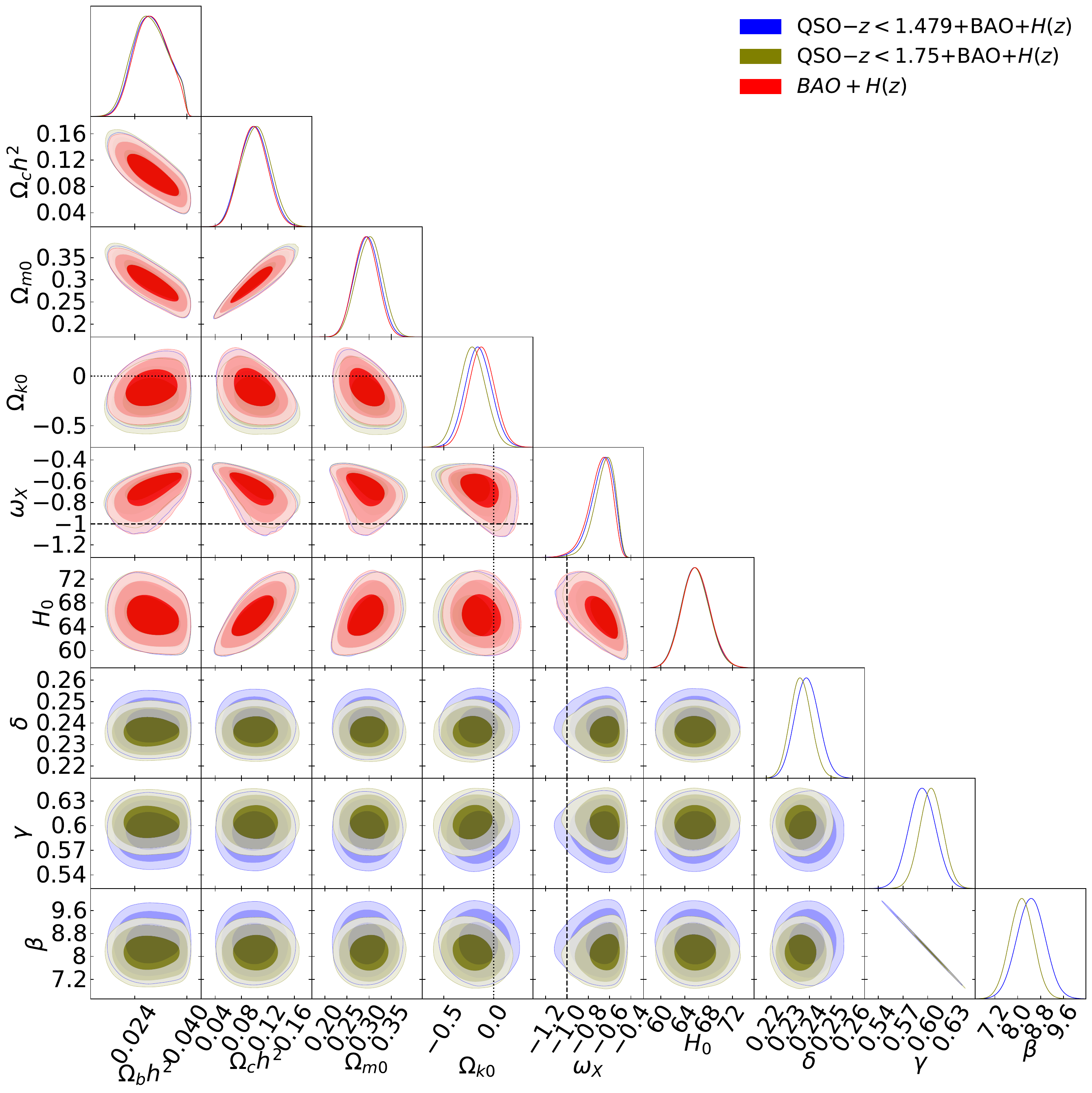}\par
\end{multicols}
\caption{One-dimensional likelihood distributions and two-dimensional contours at 1$\sigma$, 2$\sigma$, and 3$\sigma$ confidence levels using QSO$-z<1.497$ + BAO + $H(z)$ (blue), QSO$-z<1.75$ + BAO + $H(z)$ (olive), BAO + $H(z)$ (red) data for all free parameters. Left panel shows the flat XCDM parametrization. The black dashed straight lines correspond to the $\omega_X = -1$ $\Lambda$CDM model. Right panel shows the non-flat XCDM parametrization. The black dashed straight lines correspond to the $\omega_x = -1$ $\Lambda$CDM model. The black dotted straight lines correspond to $\Omega_{k0} = 0$.}
\label{fig:Eiso-Ep}
\end{figure*}

\begin{figure*}
\begin{multicols}{2}
    \includegraphics[width=\linewidth]{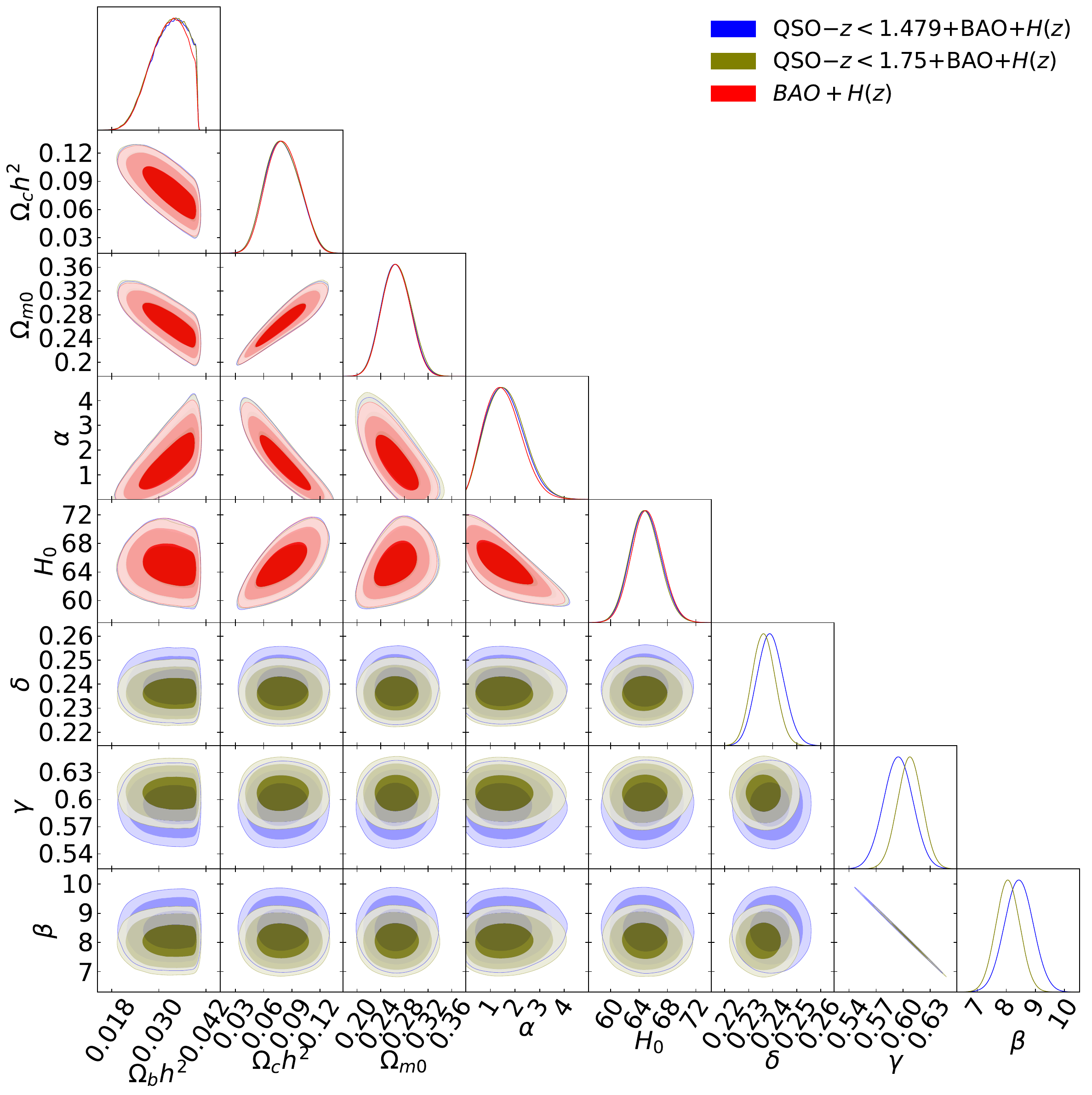}\par
    \includegraphics[width=\linewidth]{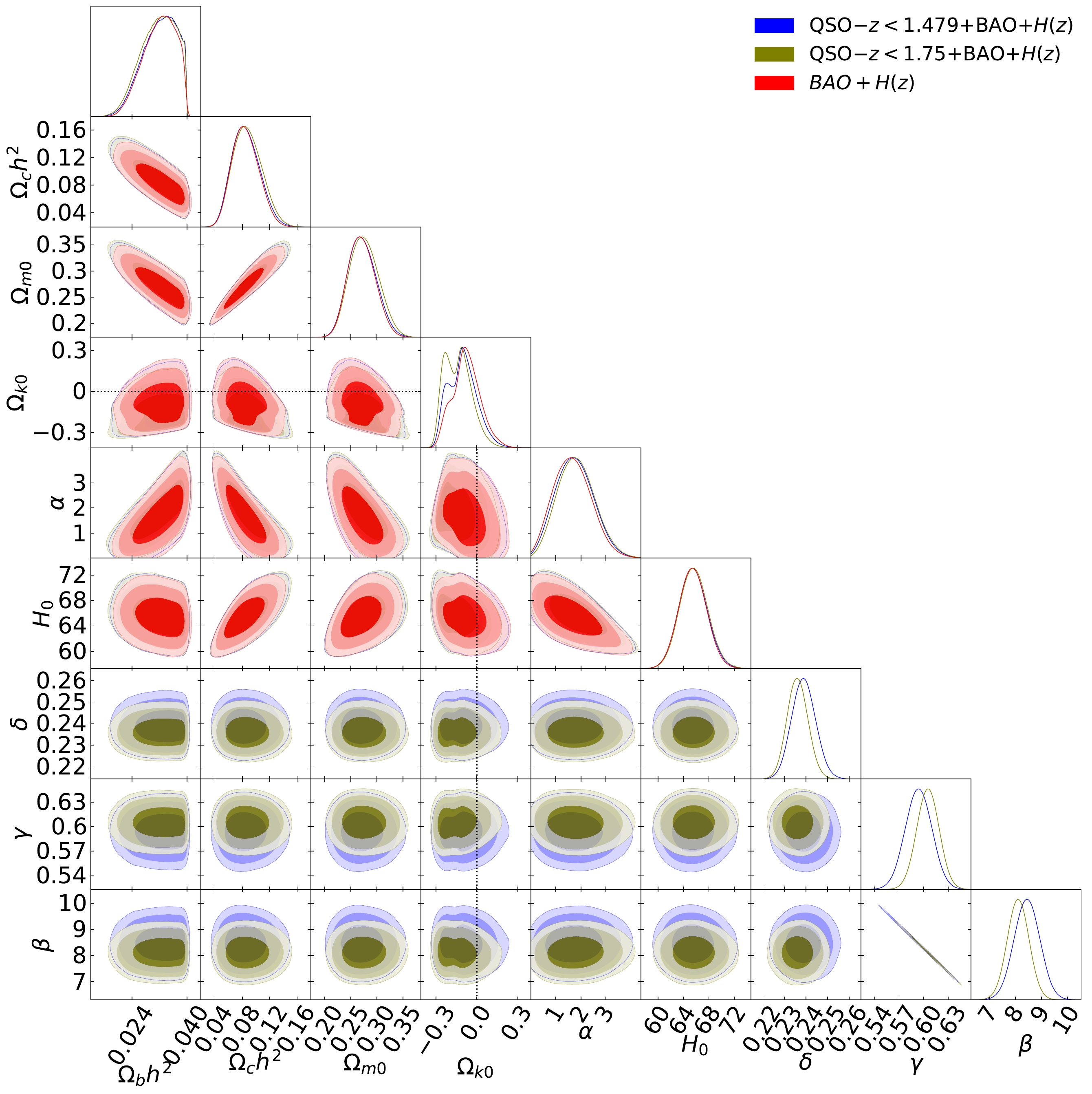}\par
\end{multicols}
\caption{One-dimensional likelihood distributions and two-dimensional contours at 1$\sigma$, 2$\sigma$, and 3$\sigma$ confidence levels using QSO$-z<1.497$ + BAO + $H(z)$ (blue), QSO$-z<1.75$ + BAO + $H(z)$ (olive), BAO + $H(z)$ (red) data for all free parameters. Left panel shows the flat $\phi$CDM model and right panel shows the non-flat $\phi$CDM model. The $\alpha = 0$ axes correspond to the $\Lambda$CDM model. The black dotted straight lines in the $\Omega_{k0}$ subpanels in the right panel correspond to $\Omega_{k0} = 0$.}
\label{fig:Eiso-Ep}
\end{figure*}

\bsp	
\label{lastpage}
\end{document}